\documentclass[12pt]{spieman}  % 12pt font required by SPIE;
\usepackage{amsmath,amsfonts,amssymb,mathtools}
\usepackage{graphicx}
\usepackage{setspace}
\usepackage{tocloft}

\title{Wavefront sensing and control in space-based coronagraph instruments using Zernike's phase-contrast method} 

\usepackage{array}
\newcolumntype{P}[1]{>{\centering\arraybackslash}p{#1}}

\author[a,*]{Garreth~Ruane}
\author[a]{J.~Kent~Wallace}
\author[a]{John~Steeves}
\author[a]{Camilo~Mejia~Prada}
\author[a]{Byoung-Joon~Seo}
\author[a]{Eduardo~Bendek}
\author[a]{Carl~Coker}
\author[a]{Pin~Chen}
\author[a]{Brendan~Crill}
\author[a]{Jeff~Jewell}
\author[a]{Brian~Kern}
\author[a]{David~Marx}
\author[a]{Phillip~K.~Poon}
\author[a]{David~Redding}
\author[a]{A~J~Eldorado~Riggs}
\author[a]{Nicholas~Siegler}
\author[a]{Robert~Zimmer}

\affil[a]{Jet Propulsion Laboratory, California Institute of Technology, 4800 Oak Grove Dr., Pasadena, CA 91109, USA}

\cftpagenumbersoff{figure}
\cftpagenumbersoff{table}

\graphicspath{{Figures/}}
\begin{document} 
  \maketitle 

%%%%%%%%%%%%%%%%%%%%%%%%%%%%%%%%%%%%%%%%%%%%%%%%%%%%%%%%%%%%% 
\begin{abstract}
Future space telescopes with coronagraph instruments will use a wavefront sensor (WFS) to measure and correct for phase errors and stabilize the stellar intensity in high-contrast images. The HabEx and LUVOIR mission concepts baseline a Zernike wavefront sensor (ZWFS), which uses Zernike's phase contrast method to convert phase in the pupil into intensity at the WFS detector. In preparation for these potential future missions, we experimentally demonstrate a ZWFS in a coronagraph instrument on the Decadal Survey Testbed in the High Contrast Imaging Testbed facility at NASA's Jet Propulsion Laboratory. We validate that the ZWFS can measure low- and mid-spatial frequency aberrations up to the control limit of the deformable mirror, with surface height sensitivity as small as 1~pm, using a configuration similar to the HabEx and LUVOIR concepts. Furthermore, we demonstrate closed-loop control, resolving an individual DM actuator, with residuals consistent with theoretical models. In addition, we predict the expected performance of a ZWFS on future space telescopes using natural starlight from a variety of spectral types. The most challenging scenarios require $\sim$1~hr of integration time to achieve picometer sensitivity. This timescale may be drastically reduced by using internal or external laser sources for sensing purposes. The experimental results and theoretical predictions presented here advance the WFS technology in the context of the next generation of space telescopes with coronagraph instruments. 
\end{abstract}

% Include a list of up to six keywords after the abstract

% Include email contact information for corresponding author
{\noindent \footnotesize\textbf{*}Address all correspondence to Garreth Ruane, Email: \linkable{garreth.ruane@jpl.nasa.gov}}\\
% % Remove this after peer-review:
% {\noindent \footnotesize\textcopyright~2020. California Institute of Technology. Government sponsorship acknowledged.}

%%%%%%%%%%%%%%%%%%%%%%%%%%%%%%%%%%%%%%%%%%%%%%%%%%%%%%%%%%%%%
\section{Introduction}
\label{sec:intro} 

A future generation of exoplanet exploration missions will aim to determine the chemical makeup of the atmospheres of a large diversity of exoplanets\cite{ESS2018}. Searching for the signatures of individual molecules in their reflectance spectra requires an instrument that can isolate the light from the planet from diffracted starlight, which would otherwise introduce noise that dominates the planet signal. High-contrast imaging with a coronagraph instrument reduces the stellar intensity at the position of the planet, thereby significantly improving the signal-to-noise ratio (S/N) for exoplanet detection and spectral characterization. 

The requirements for coronagraph instruments on future space telescopes, such as the Habitable Exoplanet Observatory (HabEx)\cite{HabEx_finalReport} and Large UV/Optical/IR Surveyor (LUVOIR)\cite{LUVOIR_finalReport} mission concepts, are driven by the goal to image and characterize Earth-like exoplanets, which appear at angular separations of $\sim$0.1$^{\prime\prime}$ and planet-to-star flux ratios of $\sim$10$^{-10}$ in the visible. To image such a planet, the coronagraph instrument creates a region of very high contrast, or ``dark hole," in the image plane with residual stellar intensity that is comparable to that of the planets of interest. The stellar intensity in the dark hole, which appears as a speckle field\cite{Soummer2007}, is determined by the precision and stability of the wavefront control system. The wavefront error tolerances are on the order of picometers at mid-spatial frequencies\cite{Ruane2018_JATIS,JuanolaParramon2019}. 

Compared to their ground-based counterparts, space-based adaptive optics (AO) systems generally need to correct smaller, slowly-varying wavefront errors to higher precision. The static wavefront correction in space-based coronagraph instruments will primarily use focal-plane wavefront sensing\cite{Giveon2011,Groff2015} and electric field conjugation (EFC)\cite{Giveon2007,Giveon2009,Pueyo2009} to determine the deformable mirror (DM) settings needed to create the dark hole. These methods have also been applied to compensate for non-common path aberrations in ground-based AO systems\cite{Potier2020}. The Coronagraph Instrument (CGI) on the upcoming Roman Space Telescope, formerly known as WFIRST, will demonstrate these methods in flight for the first time\cite{Noecker2016,Seo2018}. 

The HabEx and LUVOIR telescope concepts are specifically designed to be ultra-stable in order to accommodate the direct imaging of Earth-like exoplanets with their respective coronagraph instruments. Indeed, once the dark hole is created, it must be maintained or updated periodically over the course of observations that may take 10-1000~hr to achieve sufficient S/N in the planet image or spectrum. However, both concepts also rely on a wavefront sensor (WFS) that enables active control of the DM surfaces during observations to help relax the telescope stability requirements. For this purpose, the WFS must (1)~operate simultaneously with science observations, (2)~be in a common-path configuration, (3)~have sufficient resolution to stabilize the contrast in the dark hole at mid-spatial frequencies, and (4)~be sensitive to picometer-level wavefront errors. 

The HabEx and LUVOIR WFS design can be used simultaneously with coronagraph observations by utilizing light at wavelengths outside of the imaging passband without significantly impacting the flux at the science camera. The WFS can make use of starlight or laser sources either within the telescope or external from the spacecraft\cite{Douglas2019}. The dichroic that sends the out-of-band light to the WFS will be combined with an existing coronagraph optic allowing the WFS to measure errors that occur anywhere along the stellar beam path. Both concepts use a Zernike WFS (ZWFS) that employs Zernike's phase-contrast method\cite{Zernike1934,Zernike1955} to achieve the required resolution and sensitivity. In short, a ZWFS shifts the phase of the core of the stellar point spread function (PSF) and as a result converts phase errors into intensity variations in a pupil image at the WFS camera\cite{Dicke1975,Vorontsov2000,Bloemhof2004,Guyon2005,Fauvarque2016}. The ZWFS technique has been successfully demonstrated on ground-based adaptive optics systems\cite{Wallace2011,NDiaye2013_ZELDA,NDiaye2016_ZELDA,Vigan2019_ZELDA} and recent simulations\cite{Moore2018} and laboratory experiments\cite{Steeves2020} have achieved picometer sensitivity. The Roman Space Telescope CGI will make use of a spatially-filtered version of the ZWFS to sense and correct low order wavefront errors using light reflected off the coronagraph focal plane mask\cite{Shi2016_JATIS,Wang2016,Shi2017,Shi2018}. When used without spatial-filtering in the image plane, the spatial resolution of a ZWFS is limited by the number of detector pixels across the beam. For context, the ZWFS is also similar in principle to a point-diffraction interferometer\cite{Smartt1975}. 

An \textit{in situ} ZWFS has many benefits beyond stabilizing the wavefront during high-contrast observations. The capability to measure the wavefront with a single WFS image without any moving parts is useful for hardware diagnostics and calibrating the models used for the dark hole algorithms. For instance, the models require accurate measurements of the DM actuator locations with respect to the beam and their deflections as a function of the voltages applied. The ZWFS can also be used to periodically re-calibrate or flatten the wavefront in order to compensate for changes in optical path lengths between optics as well as drift, hysteresis, or other instabilities in the DM surface. Furthermore, the ZWFS may be used as the fine phasing sensor for a segmented primary mirror\cite{Dohlen2006,Surdej2010,JaninPotiron2017}. 

In this paper, we show that the \textit{in situ} ZWFS concept meets the sensitivity, spatial resolution, and closed-loop control requirements for directly imaging exoplanets with future space telescopes, including the HabEx and LUVOIR mission concepts. After validating our theoretical models in an experimental coronagraph instrument, we also predict the performance and limitations using natural starlight and laser illumination for wavefront sensing. In section~\ref{sec:setup}, we present the experimental setup we use to validate the ZWFS concept. In section~\ref{sec:theory}, we review the theory of the ZWFS, especially for differential wavefront measurements. In section~\ref{sec:results}, we summarize our experimental results. In section~\ref{sec:future}, we provide performance predictions for future space telescopes. Finally, we summarize our findings in section~\ref{sec:conc}.

\section{Experimental setup}\label{sec:setup}

\begin{figure}[t]
    \centering
    \includegraphics[width=\linewidth]{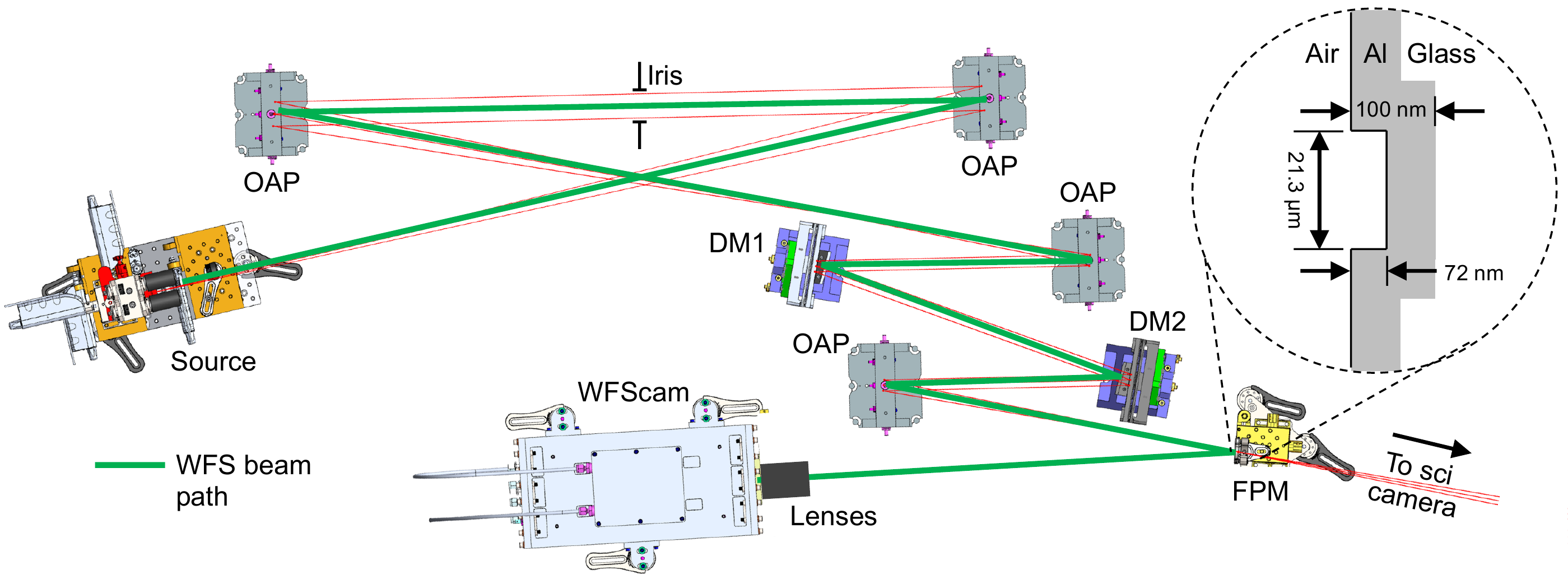}
    \caption{Schematic of the Decadal Survey Testbed (DST) wavefront sensor (WFS) beam path. The plane of the iris and DM1 are conjugate to the WFS camera (WFScam) sensor. The inset is a diagram of the WFS focal plane mask (FPM), which has a 100~nm thick aluminum coating covering a circular region 10~mm in diameter on a 1~mm thick glass substrate. A circular depression, or ``dimple," at the center of the FPM (72~nm in depth, 21.3~$\mu$m in diameter) applies the required phase shift for wavefront sensing upon reflection. %Although this aluminum mask is not designed for simultaneous use of the WFS and science camera, such capability would be made possible in a future instrument by replacing the aluminum coating with a partially reflective or dichroic coating. 
    OAP:~Off-axis parabola. DM:~Deformable mirror.}
    \label{fig:dst}
\end{figure}

We tested a prototype ZWFS in a coronagraph system in the High Contrast Imaging Testbed (HCIT) facility at NASA's Jet Propulsion Laboratory. In this section, we describe the testbed and ZWFS implementation. 

The Decadal Survey Testbed (DST)\cite{Patterson2019,DSTroadmap} is NASA's primary proving ground for the coronagraph technologies needed to image temperate, terrestrial exoplanets in reflected light with a future space telescope. The DST has demonstrated two-sided dark holes with raw contrast of $\sim$10$^{-10}$ using two DMs and a Lyot coronagraph\cite{Seo2019}. We installed and tested the ZWFS within the DST coronagraph instrument in a configuration that is similar to the HabEx and LUVOIR design. 

Figure~\ref{fig:dst} shows a schematic of the DST with the WFS beam path highlighted in green. Light from a supercontinuum laser (NKT SuperK) is focused onto a custom-made 3~$\mu$m pinhole to create a quasi-point source. A variable filter (NKT VARIA) selects the central wavelength, $\lambda_0$, and bandwidth, $\Delta\lambda$. An off-axis parabolic mirror (OAP; $f$~=~1.5~m) collimates the beam, which is truncated by an iris 47.5~mm in diameter. Two OAPs create an image of the pupil on the first DM (DM1). The light then propagates 0.6~m to the second DM (DM2) followed by an OAP that focuses the beam onto the focal plane mask (FPM). 

The DST is designed to test different sets of DMs. We carried out the following work with two experimental Boston Micromachines 2K MEMS DMs\cite{Bifano2011} with 50 actuators across the active area (47.5 illuminated). This particular set of DMs was relatively low-grade compared to those used in previous studies\cite{Seo2019} and had several actuators that were non-responsive or exhibited anomalous behavior. Parallel efforts are underway to install higher quality MEMS DMs on DST and improve their control electronics\cite{Bendek2020}. However, the set up was sufficient to demonstrate the ZWFS and characterize its performance. 

In past experiments\cite{Seo2019}, the FPM was a Lyot coronagraph mask that blocked the core of the PSF and allowed off-axis light to continue to the remainder of the coronagraph instrument and science camera. In this work, we replaced the coronagraph focal plane mask with a reflective ZWFS mask with a circular depression (or ``dimple") in an 100~nm thick aluminum coating that applied a phase shift, $\theta=4\pi h_\text{Z}/\lambda_0$, to the core of the PSF upon reflection, where $h_\text{Z}$ is the depth of the dimple (see Fig.~\ref{fig:dst}, inset). The depth and diameter of the dimple were $h_\text{Z}$~=~72~nm and $d$~=~21.3~$\mu$m, respectively, and the aluminum coating covered a circular area 10~mm in diameter on a 1~mm thick, transmissive glass substrate. The mask was provided by SILIOS Technologies. The beam focal ratio, $F^{\#}$, at the FPM was 35, which is the same as the DST coronagraph mask designs as well as for the FPMs on Roman Space Telescope CGI. The FPM alignment mechanism was a three-axis stage with motorized linear actuators.% (Newport LTAHSPPV6). 

The light reflected off of the FPM created an image of the pupil on the WFS camera (WFScam; Andor Neo sCMOS) approximately 800~mm away using two 150-mm focal length lenses (L1 \& L2; Thorlabs AC508-150-A) fixed to the front of the WFScam's custom-made enclosure. L1 \& L2 were approximately 15~mm apart and the second lens was approximately 45~mm from the sensor. The WFScam, along with the lenses, is on a motorized translation stage with 100~mm of travel that can adjust the WFScam focus. The camera position used in the experiments presented here gave an image with 660~pixels across, which can be modified by translating the camera, changing the inter-lens distance, and/or changing the lens focal lengths. The image in the WFScam was focused on DM1. Table~\ref{tab:ZWFSparams} summarizes relevant parameters for the experiments reported here. In addition, the DST was situated inside of a thermally-controlled vacuum chamber held at a pressure of $\sim$0.1~mTorr during the ZWFS testing to remove air turbulence. The WFScam and DM electronics were liquid-cooled with H$_2$O at 17$^\circ$C.  

When the beam is centered on the dimple on the reflective FPM mask, phase errors upstream of the FPM are converted into intensity variations at the WFScam. Using the analytical models developed in the next section, the wavefront errors can be estimated from the WFS images.

\begin{table}[t]
    \centering
    \begin{tabular}{|c|c|}
        \hline
        \multicolumn{2}{|l|}{\textbf{Deformable mirrors (DMs)}}\\\hline
        Number of actuators across DM & 50 \\\hline
        Number of actuators across beam & 47.5 \\\hline
        Actuator pitch & 400~$\mu$m\\\hline
        DM1-DM2 distance & 0.6~m\\\hline
        \multicolumn{2}{|l|}{\textbf{Focal plane mask (FPM)}}\\\hline
        Focal ratio at FPM, $F^\#=f/D$ & 35 \\\hline
        Reflective region diameter & 10~mm \\\hline
        Dimple depth, $h_\text{Z}$ & 72$\pm$3~nm \\\hline
        Dimple diameter, $d$ & 21.3$\pm$0.5~$\mu$m \\\hline
        Al coating thickness & 100$\pm$3~nm \\\hline
        \multicolumn{2}{|l|}{\textbf{WFS camera (WFScam)}}\\\hline
        FPM-WFScam distance & 800~mm\\\hline
        Lens (L1\&L2) focal lengths & 150~mm \\\hline 
        L1-L2 distance & 15~mm\\\hline 
        L2-sensor distance & 45~mm\\\hline 
        Beam diameter & 660~pixels \\\hline
        Pixel pitch & 6.5~$\mu$m \\\hline
        % Detector type & sCMOS \\\hline
    \end{tabular}
    \caption{DST Zernike WFS components and relevant properties. }
    \label{tab:ZWFSparams}
\end{table}

% Andor Neo 
% Interesting info on amplifiers: https://andor.oxinst.com/learning/view/article/dual-amplifier-dynamic-range
% Says 0.45 e-/ADU 
% QE = 0.6 (electrons per photon)
% https://andor.oxinst.com/learning/view/article/count-convert

\begin{figure}[t]
    \centering
    \includegraphics[width=\linewidth]{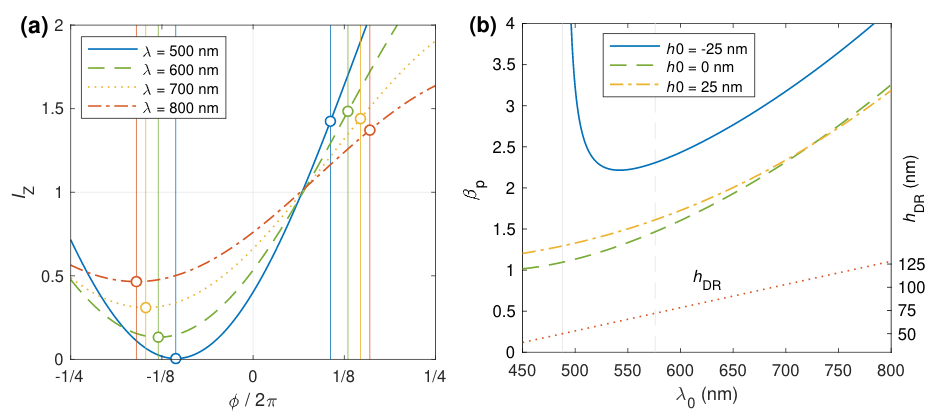}
    \caption{Theoretical performance of the DST ZWFS with dimple depth $h_\text{Z}$~=~72~nm and diameter $d$~=~21.3~$\mu$m. (a)~The ZWFS image intensity, $I_\text{Z}$, as a function of phase for different wavelengths. $I_\text{Z}$ is averaged over the pupil image and normalized by the intensity without the phase dimple, $A^2$. The markers indicate the values at the boundaries of the symmetric dynamic range, which is limited to approximately 1/4 wave by the negative bound and increases as a function of wavelength. The WFS response is defined by the slope of $I_\text{Z}$ at $\phi=0$. (b)~The WFS sensitivity, $\beta_p$, as a function of wavelength (smaller is better). The $h_0$ values show the impact of initial wavefront errors in units of surface height (i.e. half of the optical path difference). %The ZWFS is generally more sensitive to phase changes at shorter wavelengths, until the lower bound of the dynamic range; e.g. the case of $h_0$~=~-25~nm increases rapidly as the wavelength approaches $\lambda_0=4(h_\text{DR}+h_\text{Z})$~=~488~nm, where $h_\text{DR}$ is the symmetric dynamic range in units of surface height. 
    The dotted line shows the symmetric dynamic range in units of surface height, $h_\text{DR}$, as a function of wavelength with the corresponding values on the right axis. As the wavelength increases, the dynamic range increases while the WFS sensitivity becomes worse. }
    \label{fig:theory}
\end{figure}

\section{Theory}\label{sec:theory}

In this section, we review the basic theory of the ZWFS, including its design, theoretical performance, and phase reconstruction methods. 

\subsection{WFS image and response}

We denote field in the pupil as $A(x,y)e^{i\phi(x,y)}$, where $A(x,y)$ is the amplitude, $\phi(x,y)$ is the phase of interest, and $(x,y)$ are the transverse spatial coordinates. Following previous authors\cite{NDiaye2013_ZELDA}, we derive in Appendix~\ref{sec:appendixDerivation} that the pupil intensity at the WFScam sensor with the FPM dimple aligned to the center of the PSF is
\begin{equation}
    I_\text{Z} = A^2 + 2 b^2(1-\cos\theta) + 2 A b \chi(\phi),
    \label{eqn:IZ}
\end{equation}
where $b$ is the so-called reference wave and $\chi(\phi) = \sin\phi \sin\theta - \cos\phi(1-\cos\theta)$. In short, $b$ is a low-pass filtered version of $A e^{i\phi}$ containing spatial frequencies up to $\hat{d}=d/(\lambda F^\#)$ cycles across the pupil, where $d$ the FPM dimple diameter. In other words, $b$ is the field at the WFScam if the ZWFS dimple were to be replaced with an aperture of diameter $d$. In the small-aberration regime, it is sufficient to assume the $b$ is real, but in general $b$ may be complex-valued.

We define the response of the ZWFS as the slope of $I_\text{Z}$ about $\phi=0$. The derivative of $I_\text{Z}$ with respect to $\phi$ is given by $dI_\text{Z}/d\phi = 2 A b \chi^\prime(\phi)$, where $\chi^\prime(\phi) = \cos\phi \sin\theta + \sin\phi(1-\cos\theta)$. Since $\chi^\prime(0) =\sin\theta$, the maximum response is achieved with $\theta = (n+1/2)\pi$, where $n\in\mathbb{Z}$. The DST ZWFS design has a maximum $dI_\text{Z}/d\phi$ (i.e. $\theta=\pi/2$) at $\lambda_0$~=~575~nm. Figure~\ref{fig:theory}a shows $I_\text{Z}$ as a function of $\phi$ for various wavelengths across the DST operating range. 

\subsection{Dynamic range}

The WFS dynamic range is bounded about $\phi=0$ by the first roots of $dI_\text{Z}/d\phi$. A ZWFS has an asymmetric dynamic range because the bounds occur at different absolute values for positive and negative $\phi$. In practice, it is desirable to maximize the symmetric dynamic range, which is given by $\Phi = \pi - |\theta|$ for $-\pi\le\theta<\pi$. Thus, designing a ZWFS to meet a desired $\Phi$ imposes the requirement that $|\theta| \le \pi - \Phi$ (all $\theta = \theta \pm 2\pi n$ are also valid solutions). %Although the nominal design that maximizes the response (i.e. with $\theta=\pi/2$) has a dynamic range of $\pi/2$, it is possible to trade dynamic range for response by reducing $\theta$; for example, using $\lambda_0$~=~800~nm on DST extends the dynamic range by 28\% in phase and reduces the response by 9.5\%. %On the other hand, both the dynamic range and the response decrease for $\lambda_0<$~575~nm and go to zero in the limit that $\theta$ approaches $\pi$; i.e. at $\lambda_0= 4 h_\text{Z}$~=~288~nm, which is well outside of the range of wavelengths available on the DST (approximately 400-800~nm). %More precisely, with $\theta=\pi$, $I_\text{Z}$ is minimized at $\phi=0$ and increases for either a positive and negative deviation from a flat wavefront and, thus, it is not possible to recover the sign of the phase error. 
Here, we do not consider multi-wavelength\cite{Vigan2011}, phase-shifting\cite{Wallace2011}, or vector\cite{Doelman2019} ZWFS solutions that enhance the dynamic range by allowing for measurements at more than one $\theta$ value; further work is needed to make these methods compatible with coronagraph instruments. It may also be possible to reconstruct the wavefront outside of the conventional dynamic range using nonlinear methods and accurate priors\cite{Moore2018}. Without such methods, in terms of surface height, the symmetric dynamic range of the DST ZWFS, $h_\text{DR}$, varies from $h_\text{Z}$~=~72~nm at $\lambda_0$~=~575~nm to 128~nm at $\lambda_0$~=~800~nm, where the surface height is defined as half of the optical path difference. 

\subsection{Sensitivity}

% \begin{equation}
%     \Delta I = 2 A b\left[ \sin\theta(\sin\phi-\sin\phi_\text{0}) - (1-\cos\theta)(\cos\phi-\cos\phi_\text{0}) \right].
% \end{equation}

% \begin{equation}
%     \Delta I = 2 A b \Delta\phi \left[ \sin\theta \cos(\phi_\text{0}) + (1-\cos\theta)\sin(\phi_\text{0}) \right].
% \end{equation}

After creating a dark hole in the image, the ZWFS may be used to measure small changes, $\Delta\phi$, from an arbitrary, initial wavefront. For $\Delta\phi \ll $~1~radian, the phase difference between two WFS measurements is given by
\begin{equation}
    \Delta\phi = \frac{\Delta I}{2 A b \chi^\prime(\phi_0)},
    \label{eqn:deltaphi}
\end{equation}
where $\Delta I$ is the difference between the two WFS images and $\phi_0$ is the initial wavefront. Here, we use Eqn.~\ref{eqn:deltaphi} to study the sensitivity of the ZWFS in the presence of random noise. This expression may also be used for simplified, linearized reconstruction of the wavefront differences, in practice. 

The sensitivity of the WFS relates the uncertainty in the phase difference measurements to the uncertainty in the intensity difference. In units of electrons, the ideal pupil image is theoretically $I_\text{e} = \Phi_\text{star} \tau \Delta\lambda A_\text{tel} q T/N_\text{pix}$, where $\Phi_\text{star}$ is the stellar flux (photons
per unit area per unit time per unit wavelength at the
primary mirror), $\tau$ is the integration time, $\Delta\lambda$ is the spectral bandwidth of the wavefront sensor, $A_\text{tel}$ is the effective collecting area of the telescope, $q$ is the quantum efficiency of the detector, $T$ is the transmission (i.e. the ratio of flux at the WFS detector to the flux the primary mirror), and $N_\text{pix}$ is the total number of pixels within the beam in WFS image. 
We introduce intensity-fraction parameters $f_A$, $f_b$, and $f_\text{Z}$, such that $A^2 = f_A I_\text{e}$, $b^2 = f_b I_\text{e}$, and $I_\text{Z} = f_\text{Z} I_\text{e}$ to easily express these parameters in the same image units. For example, with $d=\lambda_0 F^{\#}$, $f_A \approx 1$, $f_b \approx 0.2$, and $f_\text{Z} \approx 0.5$. 

The error in the phase difference estimate due to random noise in the WFS images is given by
% \begin{equation}
%     \sigma_{\Delta\phi}^2 = \frac{\sigma_I^2}{2 A^2 b^2\beta^2} = \frac{\sigma_I^2}{2 f_A f_b I_\text{e}^2 \beta^2}, 
% \end{equation}
\begin{equation}
    \sigma_{\Delta\phi} = \frac{\sigma_I/I_\text{e}}{\sqrt{2 f_A f_b}  \chi^\prime(\phi_0) }, 
    \label{eqn:errortheory}
\end{equation}
where $\sigma_I$ is the standard deviation of the WFS image noise. In the photon-noise limit, $\sigma_I = \sqrt{f_\text{Z} I_\text{e}}$ and $\sigma_{\Delta\phi} = \beta_p / \sqrt{2 I_\text{e}}$, where $\beta_p$ is defined as the WFS sensitivity\cite{Guyon2005}:
\begin{equation}
    \beta_p = \frac{1}{\chi^\prime(\phi_0)}\sqrt{\frac{f_\text{Z}}{f_A f_b}}.
    \label{eqn:errortheory2}
\end{equation}
The raw photon counting noise in units of electrons is $\sqrt{2 I_\text{e}}$, where $2 I_\text{e}$ represents the total counts per pixel in the two images used to determine the phase difference. A smaller $\beta_p$ means the WFS is more sensitive because the uncertainty in the phase difference is smaller for a given uncertainty in the image intensity difference; an ideal ZWFS has $\beta_p = 1$. For order-of-magnitude estimates, it is convenient to assume $\beta_p=\sqrt{2}$. In that representative case, the noise in the phase difference measurement is $ \sigma_{\Delta\phi} \approx 1/\sqrt{I_\text{e}}$. In the visible regime, the wavefront sensitivity requirements are on the order of 1~pm or $\sim$10$^{-5}$ radians, which means that 10$^{10}$ photo-electrons are required per pixel in each WFS image. 

In order to analytically model the sensitivity of the DST ZWFS, we assume $f_A\approx1$ and $f_b \approx \bar{b}^2$, where $\bar{b}$ is the mean value of $b$ over a circular pupil relative to $A=1$:
% \begin{equation}
%     \bar{b} = A\left[ 1- J_0\left(\pi \hat{d}/2\right)- \frac{1}{8}\left(\pi \hat{d}/2\right)^2 J_2\left(\pi \hat{d}/2\right) \right],
%     \label{eqn:bmodel}
% \end{equation}
\begin{equation}
    \bar{b} \approx 1- J_0\left(\pi \hat{d}/2\right)- \frac{1}{8}\left(\pi \hat{d}/2\right)^2 J_2\left(\pi \hat{d}/2\right),
    \label{eqn:bmodel}
\end{equation}
where $\hat{d}=d/(\lambda F^\#)$ is the normalized dimple diameter (see Appendix~\ref{sec:appendixDerivation}). Under these assumptions, Eqn.~\ref{eqn:IZ} may be written
\begin{equation}
    f_\text{Z} = 1 + 2 \bar{b}^2 (1-\cos\theta) + 2 \bar{b} \chi(\phi_0).
    \label{eqn:fZ}
\end{equation}
Figure~\ref{fig:theory}b shows the resulting sensitivity of the DST ZWFS as a function of wavelength. The three lines in the legend (solid, dashed, and dot-dashed) are cases with different initial wavefront values, which we define in units of surface height as $h_0 = \phi_0 \lambda_0 / (4\pi)$. 
The ZWFS is generally more sensitive to phase changes at shorter wavelengths, except when the wavefront error approaches the lower bound of the dynamic range. For example, since $h_0$~=~-25~nm requires $h_\text{DR}\ge2h_0$~=~50~nm to make a differential measurement, $\beta_p$ increases rapidly as the wavelength approaches $\lambda_0=4(h_\text{DR}+h_\text{Z})$~=~488~nm, where $h_\text{DR}$~=~50~nm. The fact that the $h_0$~=~0 and $h_0$~=~25~nm cases have much better sensitivity over the whole wavelength range emphasizes the asymmetric behavior of the ZWFS. The dotted line shows $h_\text{DR}$ as a function of wavelength for comparison. To summarize Fig.~\ref{fig:theory}b, there is a fundamental trade off between WFS sensitivity and the dynamic range, but the sensitivity also degrades significantly near the lower bound of the dynamic range.

% The signal-to-noise ratio, $\Gamma$, of a phase difference measurement is defined as $\Gamma = \Delta\phi/\sigma_{\Delta\phi}$; in the photon-noise limit, $\Gamma = \Delta\phi \sqrt{2 I_\text{e}}/\beta_p$.
% % \begin{equation}
% %     \Gamma = \Delta\phi \sqrt{2 I_\text{e}}/\beta_p.
% % \end{equation}
% % \begin{equation}
% %     \Gamma = \Delta\phi \chi^\prime(\phi_0) \sqrt{2 f_A f_b I_\text{e} / f_\text{Z}}.
% % \end{equation}
% For order-of-magnitude estimates, it is convenient to assume $\beta_p=\sqrt{2}$. In that representative case, the signal-to-noise ratio of the phase difference measurement is $\Gamma \approx \Delta \phi \sqrt{I_\text{e}}$. In the visible regime, the wavefront sensitivity requirements are on the order of 1~pm or $\sim$10$^{-5}$ radians, which requires 10$^{10}$ photo-electrons per pixel per image in order to achieve $\Gamma$~=~1 in the difference measurement. 

\subsection{Impact of dimple size}

For the DST design, we chose the dimple diameter such that $\hat{d}=1.06$ at the wavelength where $\theta=\pi/2$ (i.e. $\lambda_0=8h_\text{Z}$). This splits approximately half of the beam power into the reference wave when used with a circular, unobscured pupil and leads to a WFS image, $I_\text{Z}$, that is uniform in intensity for a flat wavefront. While this configuration has practical benefits, e.g. it reduces the chances of a part of the WFS image becoming under-exposed or saturated, a larger $\hat{d}$ gives better sensitivity in theory because it increases the amplitude of $b$ in the pupil and, thereby, decreases $\beta_p$. However, this conclusion assumes that $b$ is well-known. In reality, the accuracy of the $b$ estimate also depends on $\hat{d}$ since the reference wave will be more affected by aberrations for larger $\hat{d}$. Although iterative estimation methods may help simultaneously solve for both $b$ and $\phi$, using a smaller $\hat{d}$ helps mitigate systematic errors. Further discussion of these and other systematic errors is given in section \ref{sec:systematics}.

\subsection{Absolute phase reconstruction}

So far, we have only considered using the ZWFS for differential phase measurements. While Eqn.~\ref{eqn:deltaphi} provides useful insight into the properties of the ZWFS, it is also possible to use an exact analytical method to estimate the phase over the full asymmetric dynamic range from each individual WFS image after careful calibration. In our experimental results presented below, we use the full analytical solution derived in the Appendix~\ref{sec:appendixDerivation} in all cases. Both reconstruction methods presented here are implemented as part of the open-source FALCO Matlab toolbox\cite{Riggs2018}. 

\section{Experimental results}\label{sec:results}

In this section, we present laboratory results using the ZWFS on DST. Specifically, we show example measurements of low- and mid-spatial frequency aberrations, demonstrate a range of operating wavelengths, describe the signature of DM2 (i.e. the out-of-focus DM), and compare the sensitivity of the DST ZWFS to our theoretical predictions. Building on these basic functionality tests, we demonstrate closed-loop control of an injected wavefront disturbance. Finally, we confirm that there is no significant performance degradation for wide spectral bandwidths, which are likely to be used for the WFS in future space-based coronagraphs.

\subsection{Alignment and calibration}

To align the FPM, we first find the optimum focus by moving the FPM mount in the $z$ direction until the edge of the Al-coated area is in focus on the science camera in the coronagraph instrument. Alternatively, the optimal $z$ may be determined using the shadow of the edge in the WFS image. Since the Al-coated region is circular, finding two or more points where the beam is focused on the edge allows us to estimate the ($x$,$y$,$z$) coordinates where the beam would be focused on the FPM dimple at the center. 

The WFS calibration consists of estimating the pupil amplitude, $A$, by taking a WFS image with the phase dimple offset by $\sim$1~mm with respect to the focused beam so that it reflects off the Al coating towards the WFScam. The result is an image of the pupil, $I_\text{cal}$, from which we compute the pupil amplitude estimate $A = \sqrt{I_\text{cal}}$. Finally, the reference wave, $b$, is calculated via a propagation model that estimates the far-field diffraction pattern from the dimple region given $\hat{d}$ and assuming an idealized, flat wavefront at the pupil. 

\subsection{Low-order aberrations}

\begin{figure}[t]
    \centering
    \includegraphics[width=\linewidth]{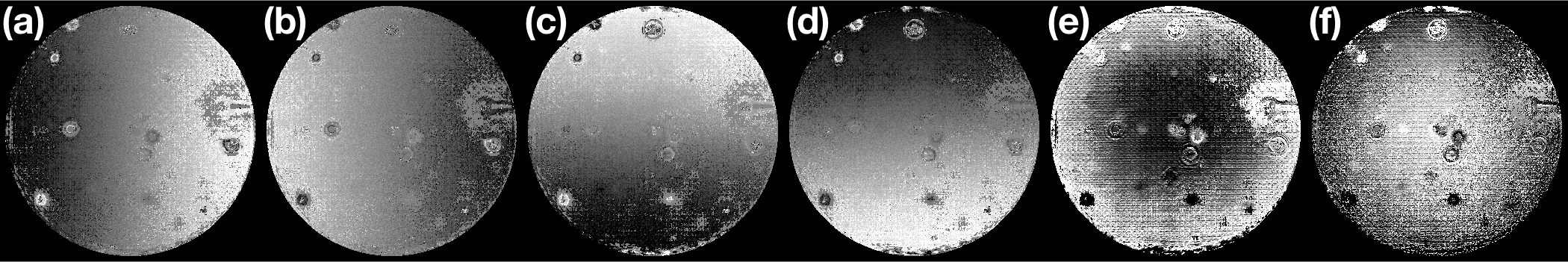}
    \caption{Difference in the wavefront measurement at $\lambda$~=~610$\pm$10~nm after moving the focal plane mask by 1~$\mu$m in the (a)~+$x$, (b)~-$x$, (c)~+$y$, (d)~-$y$ directions and 200~$\mu$m in the (e)~+$z$ and (f)~-$z$ directions. The color scale range is 28~nm, which corresponds to the theoretical peak-to-valley for (a)-(d). The expected peak-to-valley for defocus is 20~nm.}
    \label{fig:fpmscan}
\end{figure}

We introduced low-order wavefront errors using two methods: (1)~by moving the FPM in three dimensions and (2)~by introducing Zernike polynomials to the DM voltage commands. Figure~\ref{fig:fpmscan} shows the result of the three dimensional scan of the FPM offsetting by $\pm$1~$\mu$m in the transverse directions ($x$, $y$) and by $\pm$200~$\mu$m along the beam ($z$). In the case of the $x$, $y$ motions (Fig.~\ref{fig:fpmscan}a-d), the measured peak-to-valley was in good agreement with the theoretical expectation: 1~$\mu$m$/F^{\#}$~=~28~nm. Likewise, the case of the $z$ offset (Fig.~\ref{fig:fpmscan}e,f) was also in good agreement with the expected peak-to-valley of 200~$\mu$m$/(8 (F^{\#})^2)$~=~20~nm. These measurements were taken with $\lambda$~=~610$\pm$10~nm and the DM in quasi-flat state with $\sim$30~nm RMS wavefront error. The anomalous actuators visible in Fig.~\ref{fig:fpmscan} were outliers in surface height and fell outside of the WFS dynamic range. We attributed the high-spatial frequency noise in Fig.~\ref{fig:fpmscan}e,f to small lateral beam shifts that occur when moving the FPM stage in the $z$ direction. Overall, the WFS responded as expected to controlled motions of the FPM. 

\begin{figure}[t]
    \centering
    \includegraphics[width=\linewidth]{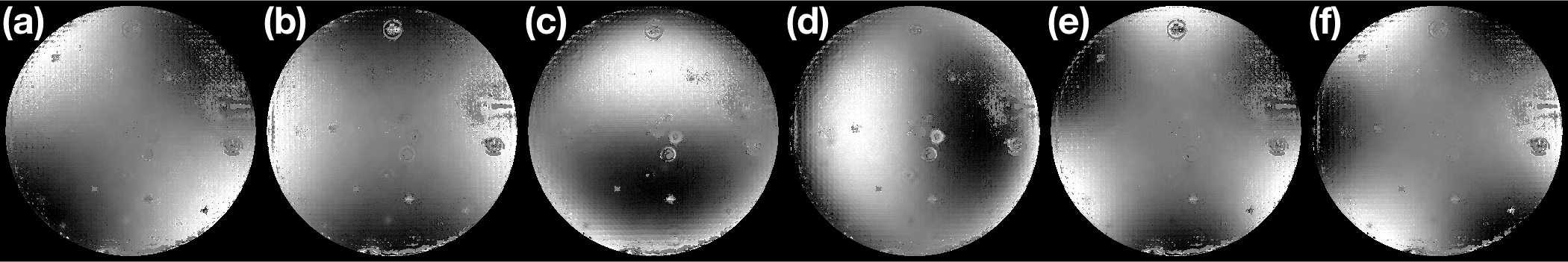}
    \caption{Difference in the wavefront measurement at $\lambda$~=~610$\pm$10~nm after applying Zernike polynomials in the DM voltage commands corresponding to (a)-(b)~astigmatism, (c)-(d)~coma, and (e)-(f)~trefoil. The color scale range is 2~nm and the change in the DM surface height in all cases is $\sim$0.3~nm RMS.}
    \label{fig:pokeZernikes}
\end{figure}

Figure~\ref{fig:pokeZernikes} shows a similar experiment, but now using DM1 to produce low-order aberrations. In this case, we 
measured the change in wavefront after adding offsets to the DM voltage commands in the form of Zernike polynomials. The resulting DM surface is not exactly a Zernike polynomial, since we did not normalize by the non-uniform actuator surface deflection per volt (also known as the actuator gains). Nonetheless, the examples in Fig.~\ref{fig:pokeZernikes} represent astigmatism, coma, and trefoil aberrations with RMS voltage equivalent to 8~bits, which resulted in a measured RMS surface height of 0.3~nm. Thus, the empirically determined conversion between the RMS DM commands and RMS surface displacement for the low order modes shown was 36$\pm$1~pm per bit, where a single bit in the DM electronics corresponds to a voltage change of $\sim$1~mV. 

When we used the DM to introduce a wavefront difference, we took the difference between positive and negative offsets added to the nominal voltage commands. For example, in this case, we applied $\pm$4~bits RMS of each Zernike polynomial to obtain a net difference of 8~bits RMS. This method is applied for the cases described below as well. 

\subsection{Mid-spatial frequency aberrations }

\begin{figure}[t]
    \centering
    \includegraphics[width=\linewidth]{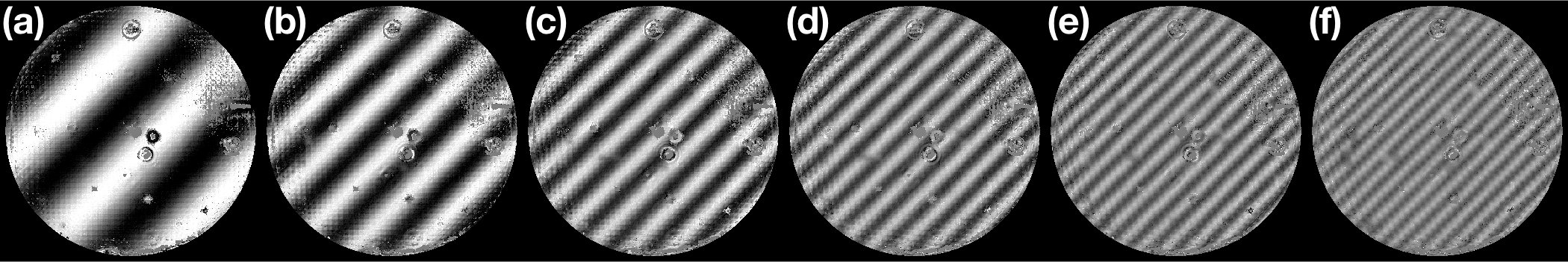}
    \caption{Same as Fig.~\ref{fig:pokeZernikes}, but with a sinusoidal voltage pattern with (a)~3, (b)~6, (c)~9, (d)~12, (e)~15, and (f)~18 cycles across the DM. The color scale range is 1~nm.}
    \label{fig:pokeSines}
\end{figure}

The DST ZWFS was specifically designed to measure mid-spatial frequency aberrations, in addition to the low-order examples shown above. Figure~\ref{fig:pokeSines} shows the measured wavefront differences after applying a sinusoidal voltage pattern with a peak-to-valley of 16~bits at spatial frequencies ranging from 3 to 18 cycles across DM1. This confirms that the ZWFS can accurately measure aberrations across a large range of spatial frequencies. The apparent reduction in the peak-to-valley of the sinusoidal aberration as a function of spatial frequency is a known property of DMs with continuous surfaces; the surface motion is usually a few times larger for low spatial frequencies compared to isolated actuator pokes for a given voltage. 

\begin{figure}[t]
    \centering
    \includegraphics[width=0.8\linewidth]{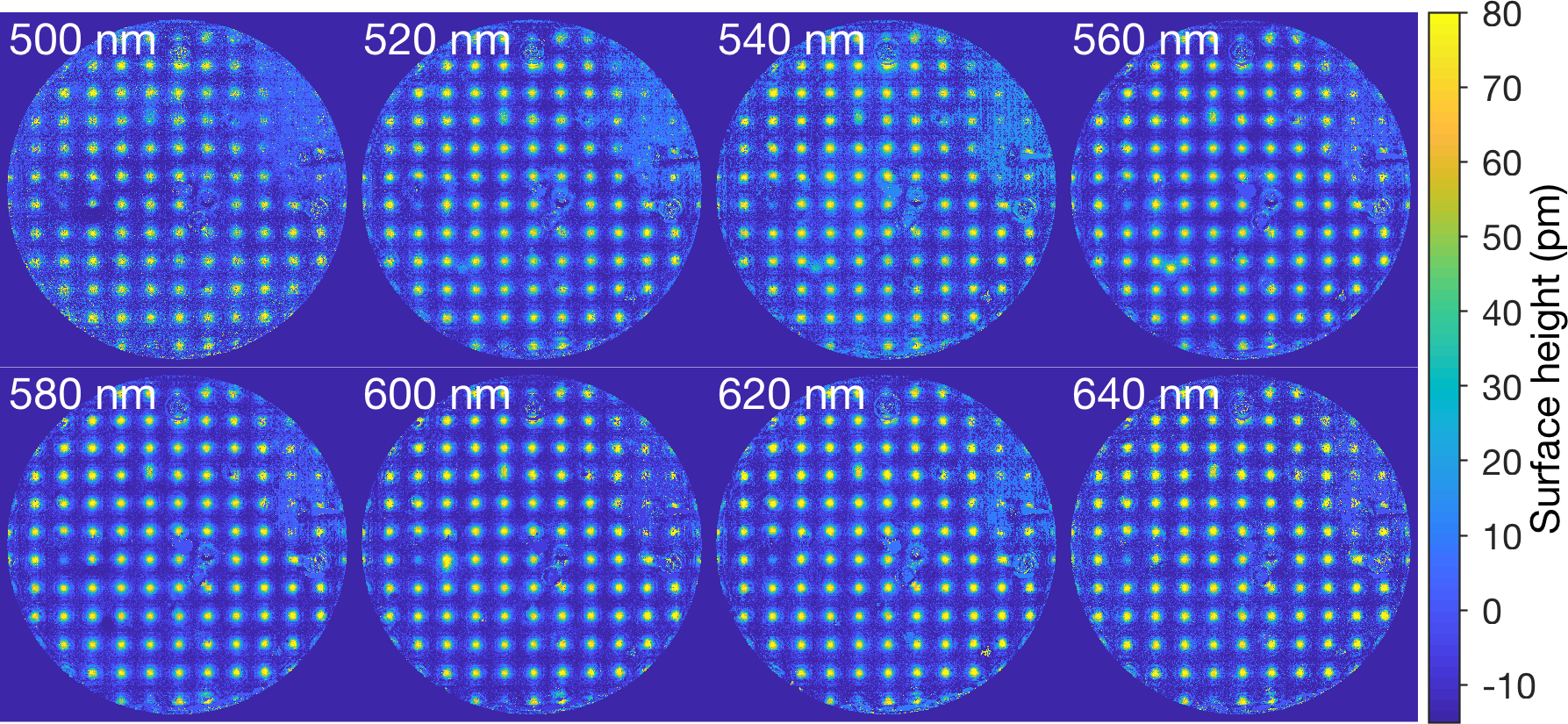}
    \caption{Measured DM surface height difference using a central wavelength ranging from 500~nm to 640~nm. The voltage of the grid of actuators was changed by 8 bits. The result is largely insensitive to the central wavelength. Some known rogue actuators are visible that randomly change the local surface height .}
    \label{fig:pokeGrid_chromaticity}
\end{figure}

To study the case of isolated actuator pokes, we offset the DM command of 1 out of every 16 actuators by 8 bits resulting in a poke-grid pattern. Figure~\ref{fig:pokeGrid_chromaticity} shows the estimated surface height after applying the poke grid, using a range of central wavelengths from 500~nm to 640~nm. In each case, the source bandwidth was 20~nm. We found that the WFS resolves the individual actuators as expected, the measured surface heights agree to within 10\%, and that there is no significant correlation with wavelength. This confirms that the DST ZWFS can be readily used to measure full range of correctable spatial frequencies over most, if not all, of the visible range. In fact, the DST ZWFS revealed that several actuators on DM1 were misbehaving; uncommanded actuator motions appear is a few cases in Fig.~\ref{fig:pokeGrid_chromaticity}. We also used these measurements to empirically determine that the conversion between the DM commands and the displacement of an isolated actuator was 11.5$\pm$0.5~pm per bit, on average. This measured value is consistent with our expectation.

\begin{figure}[t]
    \centering
    \includegraphics[width=0.5\linewidth]{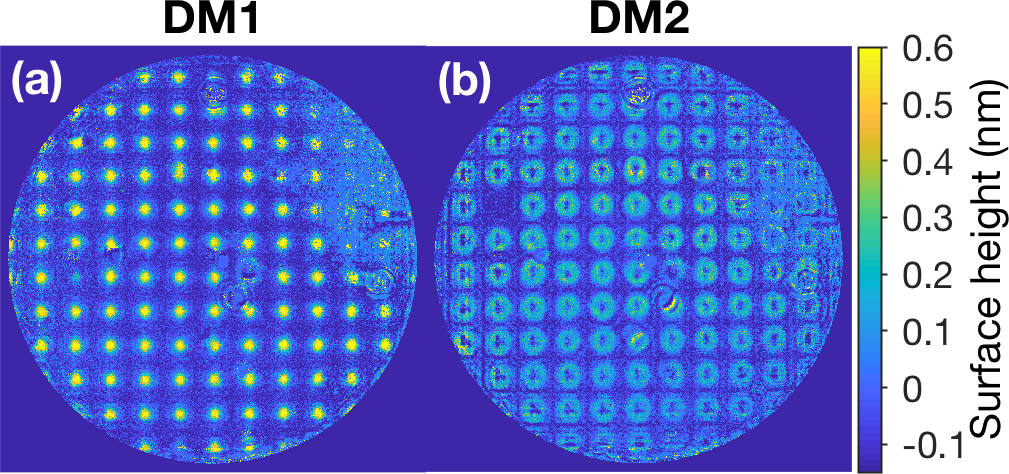}
    \caption{Measured DM surface height difference using a central wavelength of $\lambda$~=~610$\pm$10~nm after applying voltage to a grid of actuators on (a)~DM1 and (b)~DM2. A single isolated actuator on DM2 appears as a ``donut" shape in the WFS measurements due to diffraction. This effect is less apparent at lower spatial frequencies. %Since DM2 modulates both amplitude and phase of the field at the pupil, 
    }
    \label{fig:pokeGrid_DM1_vs_DM2}
\end{figure}

\subsection{Diffraction from out-of-plane optics}

Since the ZWFS is in a common-path configuration, the measured phase is the result of propagation through all of the upstream optics, including DM1 and DM2. In the examples above, we have only applied commands to DM1, which is in-focus in the ZWFS image. However, DM2 is 0.6~m from DM1 and, therefore, significant diffraction occurs as the light propagates from DM2 to the WFS image plane (the Fresnel number is 231 for $\lambda$~=~610~nm). In fact, the diffraction between DM1 and DM2 is an important design feature for creating two-sided dark holes with the coronagraph\cite{Pueyo2009}. 

Figure \ref{fig:pokeGrid_DM1_vs_DM2} compares the ZWFS measurement with a poke grid pattern applied to DM1 versus DM2. The actuators on DM2 appear as ``donut" shapes due to the propagation distance between the DMs. Moving an actuator on DM2 modulates both the amplitude and phase at the pupil plane that is measured by the ZWFS. Therefore, the calibration of $A$ is no longer valid after applying significant surface features to DM2. This effect can limit the ability of the ZWFS to accurately estimate the wavefront when errors are introduced at optics that are not conjugate to the pupil plane. We refer to such optics as ``out-of-plane." 

When using the ZWFS to calibrate the DM models, we match the ZWFS measurements of both DM1 and DM2 to a simulation of the ZWFS image that includes their respective propagation distances. This allows us to accurately determine the position of the actuators with respect to the beam, the relationship between the DM commands and the displacement of the actuators, and which actuators are unresponsive or exhibit anomalous behavior. Thus, the propagation does not negatively impact our ability to use the ZWFS as a DM calibration tool. 

The diffraction from out-of-plane optics may limit the wavefront control performance when using the ZWFS measurements to stabilize the wavefront in the coronagraph instrument. On one hand, the aberrations generated at the primary mirror of the telescope are practically in the pupil plane, they can be accurately sensed by the ZWFS and readily corrected with DM1. On the other hand, some aberrations generated at out-of-plane optics may not be accurately sensed, especially if they have significant mid-spatial frequency content, which causes more diffraction to occur compared to low spatial frequencies. A common culprit for such errors is a lateral motion of the beam across the surface of out-of-plane optics, also known as beam walk\cite{Noecker2005}. This effect, and the challenges associated with measuring it, must be taken into account when setting the wavefront stability requirements for future space telescopes.

\subsection{Experimental sensitivity measurement}

\begin{figure}[t]
    \centering
    \includegraphics[width=\linewidth]{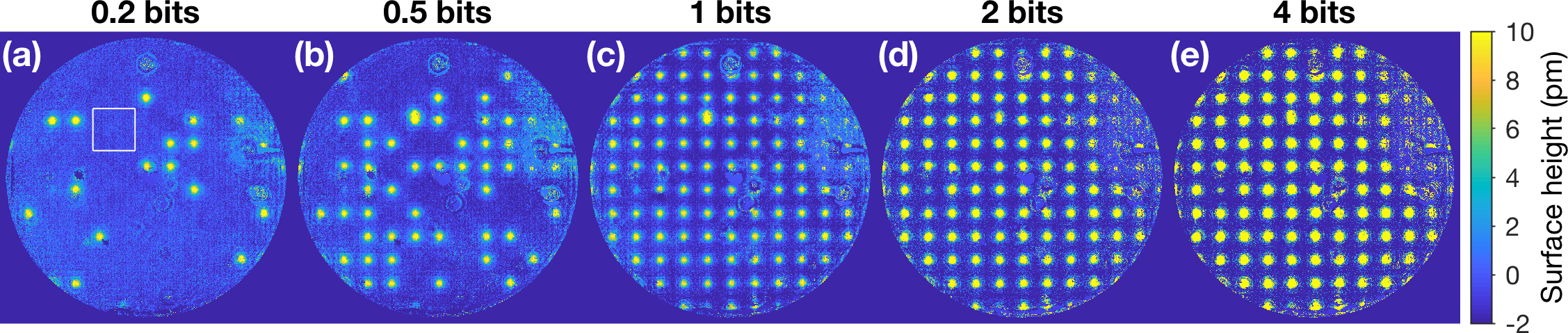}
    \caption{Measured DM surface height difference after changing the voltage of a grid of actuators by (a)~0.2, (b)~0.5, (c)~1, (d)~2, and (e)~4 bits. (a) and (b) show that a proportional number of actuators respond when the command is less than 1 bit, which confirms that the DM actuators were moving by the minimum possible height change ($\sim$10~pm) corresponding to the least significant bit of the DM electronics. This result was obtained after combining 2$\times$10$^6$ WFS frames. The noise floor, determined from the standard deviation of the values in the 100$\times$100 pixel white square, was $<$1~pm. The wavelength was $\lambda$~=~610$\pm$10~nm.}
    \label{fig:pokeGridSens}
\end{figure}

In order to compare the sensitivity of the DST ZWFS to theoretical predictions, we performed an experiment where we introduced the grid of actuator pokes shown in Figs.~\ref{fig:pokeGrid_chromaticity} and \ref{fig:pokeGrid_DM1_vs_DM2} on DM1 and reduced the voltage offsets until we reached a single-bit difference. To confirm that we were in fact moving the actuators by one bit, we applied fractional (0.5 and 0.2) bit offsets to the voltage commands prior to quantization and found that the expected fraction of actuators responded since the initial actuator voltages were approximately uniformly-distributed over a large portion of the full voltage range. The fact that the surface height displacement of the triggered actuators was the same in the fractional bit cases as the one-bit case provided further confirmation (see Fig.~\ref{fig:pokeGridSens}). 

\begin{figure}[t]
    \centering
    \includegraphics[height=0.45\linewidth]{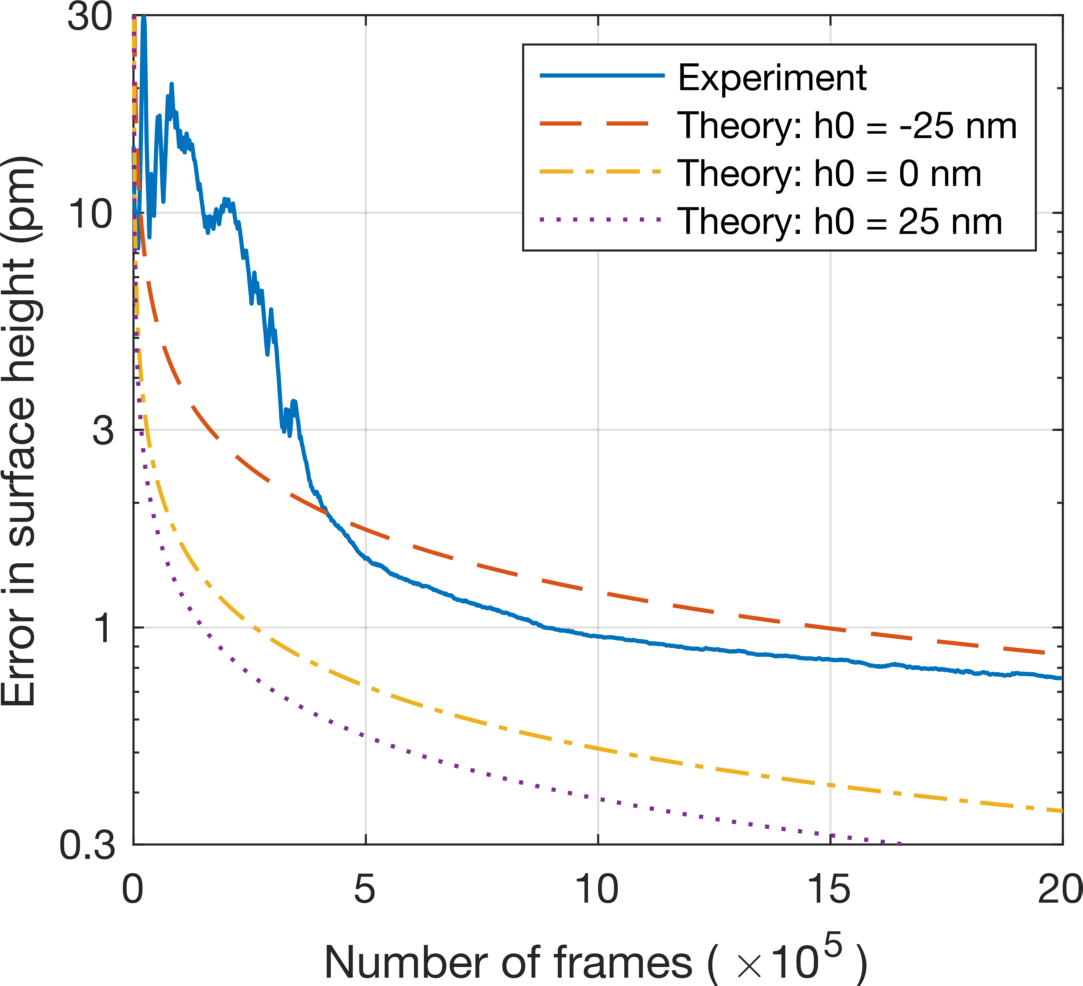}
    \caption{Comparison between the empirical error in the surface height difference measurements and the theoretical predictions for different initial surface heights. The empirical error is defined as the standard deviation within the 100$\times$100 pixel white square in Fig.~\ref{fig:pokeGridSens}a.}
    \label{fig:dst_eb}
\end{figure}

For each case in Fig.~\ref{fig:pokeGridSens}, we switched back and forth between the two DM states every 1000~frames, which were taken at 60~frames per second with discrete integration times of 10~ms per frame. We repeated this for 1000~cycles resulting in 2$\times$10$^6$ frames. To combine the data, we took the median of the difference measurements. Figure~\ref{fig:dst_eb} shows the resulting surface height error (defined as the standard deviation within the 100$\times$100 pixel white square in Fig.~\ref{fig:pokeGridSens}a), versus the number of combined frames. The noise floor reached 1~pm after combining 4.4$\times$10$^5$ frames with an effective integration time of 1.2~hr per DM state (2.4~hr total). 

To compare with the theoretical predictions in Section \ref{sec:theory}, we estimated $I_\text{e}$ from the empirical count rate, $\dot{N}_\text{c}$ (ADU per sec per pixel), in the pupil image with the phase dimple misaligned. The pupil image in units of electrons is $I_\text{e} = n_\text{f} \tau_\text{f} G \dot{N}_\text{c}$, where $n_\text{f}$ is the number of frames, $\tau_\text{f}$ is the discrete integration time per frame, and $G$ is the detector gain (e$^{-}$ per ADU). Including the dark current and read noise of our sCMOS detector, the noise variance in the WFS images becomes
\begin{equation}
    \sigma_I^2 = n_\text{f} \left(f_\text{Z} G \dot{N}_\text{c}\tau_\text{f} + i_\text{d} \tau_\text{f} + N_\text{read}^2 \right),
\end{equation}
where $i_\text{d}$ is the dark current (e$^{-}$ per second) and $N_\text{read}$ is the read noise. Using the experimental parameters in Table~\ref{tab:ExperimentParams}, we determined the theoretical noise floor by plugging the predicted $\sigma_I$ into Eqn.~\ref{eqn:errortheory}. The dashed, dot-dashed, and dotted lines in Fig.~\ref{fig:dst_eb} show the theoretical error in the surface height difference measurements for three initial surface heights: $h_0$~=~-25~nm, $h_0$~=~0~nm, and $h_0$~=~25~nm. 

\begin{table}[h]
    \centering
    \begin{tabular}{|c|c|c|}
        \hline
        Name & Symbol & Value \\\hline
        Central wavelength & $\lambda_0$ & 610~nm \\\hline
        Spectral bandwidth & $\Delta\lambda$ & 20~nm \\\hline
        Dimple phase shift & $\theta$ & 1.48~radians \\\hline
        Dimple diameter & $d$ & 1.0~$\lambda_0 F^\#$ \\\hline
        Count rate$^*$ & $\dot{N}_\text{c}$ &
        2.6$\times$10$^{6}$ ADU/sec \\\hline
        Pupil intensity frac.$^*$ & $f_\text{A}$ & 1.0 \\\hline
        Ref. wave intensity frac.$^*$ & $f_\text{b}$ & 0.20 \\\hline
        WFS image intensity frac.$^*$ & $f_\text{Z}$ & 0.53 \\\hline
        Discrete integration time & $\tau_\text{f}$ & 10~ms/frame\\\hline
        Quantum efficiency & $q$ & 0.6 \\\hline
        Read noise (rms) & $N_\text{read}$ & 2~e$^{-}$ \\\hline
        Dark current & $i_\text{d}$ & 0.01~e$^{-}$/pixel/sec \\\hline
        % Pixel well depth & 30,000~e$^{-}$ \\\hline
        Detector gain & $G$ & 0.45 e$^{-}$/ADU \\\hline
        
    \end{tabular}
    \caption{DST Zernike WFS experimental parameters for the sensitivity measurement. $^*$Median of across the pupil. $F^\#$ is the focal ratio at the focal plane mask.}
    \label{tab:ExperimentParams}
\end{table}

Compared to the case of a perfectly flat wavefront (i.e. $h_0$~=~0~nm), the experimental error was within a factor of 2 of the theoretical value after combining 5$\times$10$^5$ frames. Since the local peak-to-valley surface error of the DM within the white square in Fig.~\ref{fig:dst_eb} was approximately 28~nm (5.2~nm~RMS), we expected the actual performance to be somewhat degraded with respect to the $h_0$~=~0~nm case. Indeed, the sensitivity of the WFS varies across the image and the theoretical curves in Fig.~\ref{fig:dst_eb} will bound typical cases. We attributed the relatively large errors in the experimental measurements on the left side of Fig.~\ref{fig:dst_eb} to testbed instabilities and tentatively concluded that the dominant instability is random beam motion with respect to the WFScam. Using a more stable FPM and WFScam mounting scheme and minimizing temperature changes during our experiments may have improved the sensitivity. Other potentially significant error sources that were not included in the theoretical predictions were excess detector noise and uncommanded DM surface motions, which were both present in the experimental data. %These sources may have also prevented the errors from reducing as $1/\sqrt{n_\text{f}}$ at larger number of frames 

\subsection{Closed-loop wavefront control demonstration}

To enable closed-loop wavefront control, we fit a model of the DM to a poke-grid calibration measurement (see e.g. Fig~\ref{fig:pokeGrid_DM1_vs_DM2}) in order to determine the location of the actuators with respect to the beam and the conversion between voltage and surface height for each actuator. The DM model assumes a linear superposition of actuator influence functions; specifically, we used the BMC 2K DM model provided with FALCO\cite{Riggs2018}.

\begin{figure}[t]
    \centering
    \includegraphics[width=0.5\linewidth]{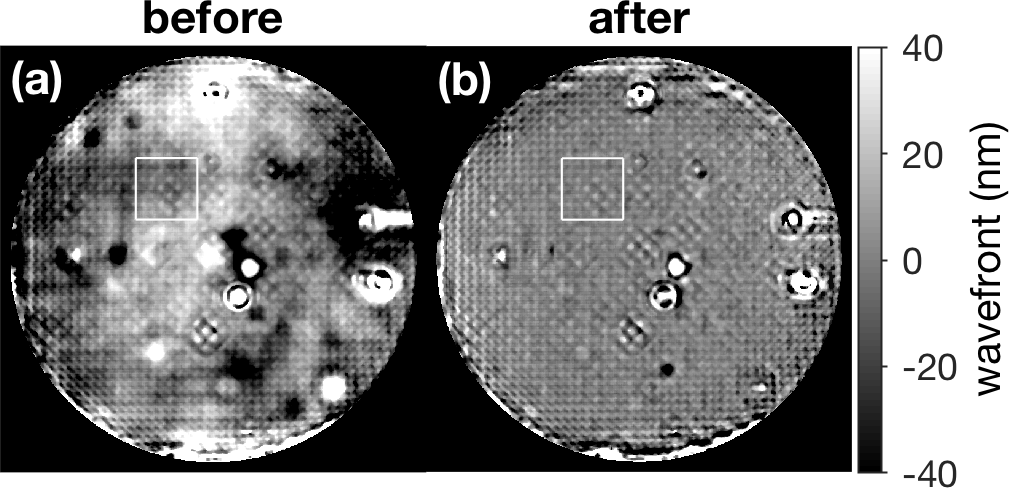}
    \caption{Closed loop flattening of the wavefront in units of the optical path difference. The wavefront error was reduced from (a)~27~nm~RMS to (b)~15~nm~RMS over the full beam including the local defects in the DM surface. Within the white square, the wavefront error is (a)~5.2~nm~RMS and (b)~2.2~nm~RMS, respectively.}
    \label{fig:flattenDM}
\end{figure}

Figure~\ref{fig:flattenDM} shows the closed-loop flattening of the wavefront by compensating for the measured phase error with DM1. After a few iterations, the wavefront error converged from 27~nm~RMS to 15~nm~RMS measured over the full DM. The reported error includes uncontrollable local defects visible on the DM surface. For comparison, the wavefront error reduced from 5.2~nm~RMS to 2.2~nm~RMS within the small white square, which covers a relatively clean region of the DM with no defects or anomalies. For these closed-loop demonstrations, we used $\lambda$~=~575$\pm$10~nm, which gives $\theta \approx \pi/2$.

\begin{figure}[t]
    \centering
    \includegraphics[width=0.75\linewidth]{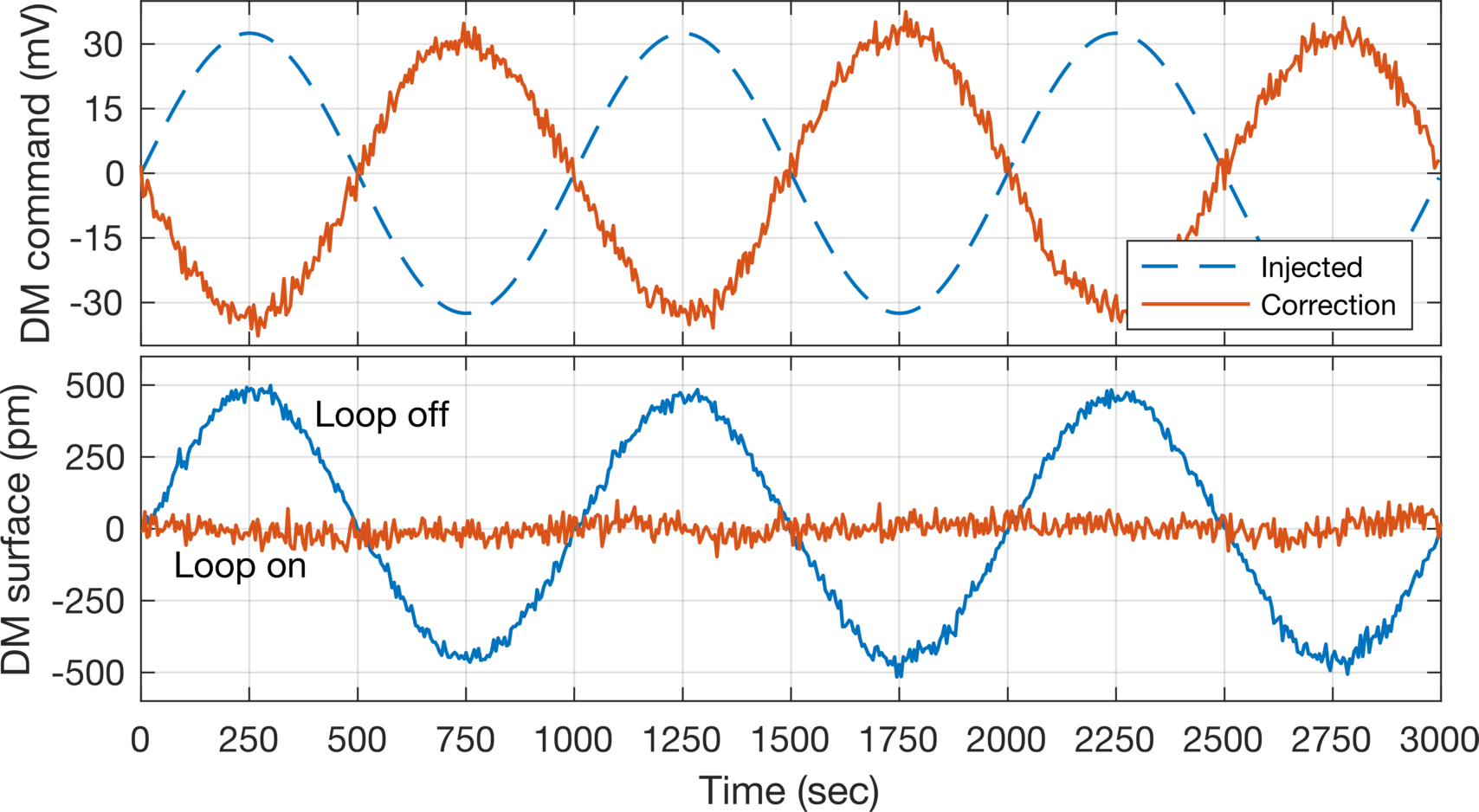}
    \caption{Demonstration of closed-loop control of a single DM actuator. (\textit{top panel}) The injected DM command (dashed line) and the correction applied based on the ZWFS measurements (solid line). (\textit{bottom panel}) Comparison between the DM surface measurements with and without the closed-loop control. The residual surface error was 34~pm~RMS for 1~sec effective integration time per measurement. }
    \label{fig:ClosedLoopDemo}
\end{figure}

One of the primary applications of the ZWFS for future space telescopes is to compensate for instabilities in the DM surface. For instance, as mentioned above, the DM electronics that we used for all of the experiments presented here exhibited significant noise and a few actuators were randomly changing by up to 1~nm in surface height. While we know from experience that uncommanded motions can be solved by tracking down faults in the DM electronics, and that DM stability on the order of a picometer is possible\cite{Seo2019}, having the capability to stabilize the DM in closed-loop can help relax the long-term DM stability requirements. To demonstrate this capability, we set up an experiment (see Fig.~\ref{fig:ClosedLoopDemo}) where we introduced a sinusoidal voltage versus time to a single DM actuator with peak-to-valley of 65~mV and period of 1000~sec (Fig.~\ref{fig:ClosedLoopDemo}, \textit{top panel}, dashed line). We then used the ZWFS measurements to correct for the injected actuator motion (Fig.~\ref{fig:ClosedLoopDemo}, \textit{top panel}, solid line) using an integration time of 1~sec per measurement (combining 100 frames with a discrete integration time of 10~ms per frame).  

The \textit{bottom panel} of Fig.~\ref{fig:ClosedLoopDemo} shows the DM surface measurements with and without the closed-loop control. By fitting a sine wave to the ``Loop off'' case, we found that the peak-to-valley in the DM surface over time was 910~pm, which indicates that the actuator moves by approximately 14~pm/mV. The ``Loop on'' case shows that the injected sinusoidal motion was removed when closed-loop control was activated. The residual surface error was 34~pm~RMS, which is consistent with our expectations based on the analytical noise model presented in the previous section. Based on the assumption that the error is proportional to $1/\sqrt{n_\text{f}}$, a similar demonstration with 1~pm~RMS residuals is feasible with (34)$^2$~=~1156 times the number of frames per measurement, which is equivalent to a WFS integration time of $\sim$20~min and the timescale for the wavefront variations would need to be increased accordingly. 

\subsection{Broadband demonstration}

\begin{figure}[t]
    \centering
    \includegraphics[width=0.95\linewidth]{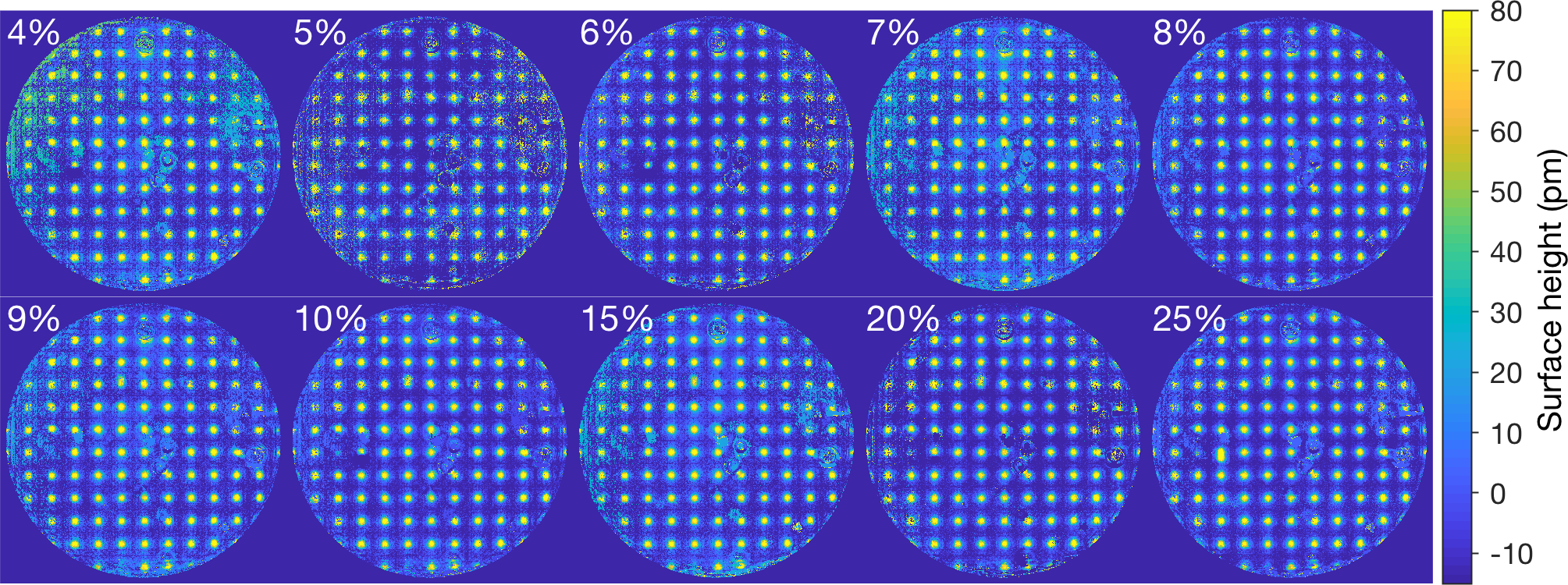}
    \caption{Measured DM surface height difference using a central wavelength of 600~nm with an increasing bandwidth (i.e. $\Delta\lambda/\lambda$ ranging from 4\% to 25\%). The voltage of the grid of actuators was changed by 8 bits. The result is largely insensitive to the bandwidth.}
    \label{fig:pokeGrid_chromaticityBW}
\end{figure}

While all of the experiments to this point used a relatively narrow spectral bandwidth of 20~nm, future space telescopes will potentially use a much larger passband (fractional bandwidths of $\Delta\lambda$ = 20\% or more) to maximize the stellar flux used by the ZWFS. With this in mind, we performed an experiment where we repeated the same poke-grid measurement using an increasing spectral bandwidth ranging from 4\% to 25\% (see Fig.~\ref{fig:pokeGrid_chromaticityBW}). We centered the bands around $\lambda_0$~=~600~nm because that optimized the broadband flux on our testbed. We did not observe any obvious degradation or error correlated with bandwidth. In fact, we found that using a larger bandwidth may help mitigate the impact of sub-actuator features, which diffract significantly when they are not perfectly in-focus and are likely not well sampled. The resulting phase errors for different wavelengths partially average out. 

An important caveat is that accurate wavefront reconstruction requires an assumption on the input spectrum. A discrepancy between the actual and assumed source spectrum can introduce systematic measurement errors. We assumed here that the input spectrum was uniform, which is reasonably accurate in the case of the DST source. However, generally speaking, optimal ZWFS performance will be achieved after calibrating the instrument transmission versus wavelength and including prior knowledge of the source spectrum to avoid biases in the wavefront measurements. 

\section{Performance predictions for future space telescopes}\label{sec:future}

Enabling closed-loop control of the wavefront in a space-based coronagraph instrument is central to the motivation for this work. However, there are several possible implementation pathways for new mission concepts, such as HabEx and LUVOIR. In this section, we briefly introduce the possible implementations of the ZWFS, calculate the expected timescales for picometer control using natural starlight or laser illumination, and provide an example wavefront sensing error budget including both random and systematic errors. 

\subsection{Potential implementations}

The favored approach for the HabEx and LUVOIR concepts is to use out-of-band starlight for wavefront sensing taken from the coronagraph beam path\cite{HabEx_finalReport,LUVOIR_finalReport}. For example, the HabEx design specifies that the focal plane substrate will have a dichroic coating to reflect the desired passband to the ZWFS, rather than reflecting all of the light as in our experimental demonstration above. The reasoning for using this location for the ZWFS pick-off is that it is only one optic away from the location where the in-band starlight is blocked by the Lyot stop, which maximizes the common path. The HabEx ZWFS can therefore measure the aberrations anywhere in upstream optical system, especially the optics that are dominant sources of wavefront instabilities: the telescope assembly and the DMs. %Alternate designs that introduce additional beamsplitters before the focal plane mask in the coronagraph are likely not possible because the coatings will have extreme requirements in order to prevent ghost reflections from spoiling the contrast in the dark hole. A viable alternative is to use a beamsplitter at or after the Lyot stop where the in-band starlight is mostly extinguished, but this requires a non-trivial wavefront reconstruction with partial wavefront information. 

Much like in ground-based AO systems, the WFS in space-based coronagraphs can either use a natural guide star (NGS) or laser illumination. The latter can be from an internal source, e.g. within the coronagraph instrument, or an external laser guide star (LGS) in a separate spacecraft\cite{Douglas2019}. The HabEx ZWFS design described above, and similar ZWFS implementations, can make use of the light from any of these types of sources. The advantage of NGS and external LGS, compared to an internal source, is that they can sense the entire beam path from the primary mirror to the ZWFS pick off. On the other hand, while the LGS can provide far more flux than NGS, it requires a separate spacecraft with formation flying capability. 

\subsection{Notional WFS requirements}

The wavefront stability requirements for a space-based coronagraph depend on the telescope and instrument design as well as the mission observing strategies and the targets of interest. Previous studies have arrived at wavefront stability requirements of 1-10~pm to observe an Earth-sized planet at 1~au from its host star\cite{Ruane2018_JATIS,JuanolaParramon2019}, which corresponds to a planet-to-star flux ratio of $\sim$10$^{-10}$. However, the broader science case includes temperate, rocky planets orbiting a wide range of stellar types. 

For a constant incident flux on the planet, the orbital separation from the star must scale with $\sqrt{L_\text{star}}$, where $L_\text{star}$ is the stellar luminosity. Late-type (K and M) stars are smaller, cooler, and less luminous than our Sun (a G star). %Since $L_\text{star}\propto R_\text{star}^2 T_\text{star}^4$, where $R_\text{star}$ and $T_\text{star}$ are the stellar radius and temperature, late-type stars are also far less luminous. 
For example, the Earth equivalent insolation distance (EEID) for an M~star is $a_\text{EEID}\approx$~0.05~au. Likewise, early-type (A and F) stars are more luminous and, thus, $a_\text{EEID}>$~1~au. The planet-to-star flux ratio, $\epsilon$, of an exoEarth may be approximated by
\begin{equation}
    \epsilon \approx 10^{-10}\left(\frac{\text{1 au}}{a_\text{EEID}}\right)^2,
\end{equation}
which ranges from $\sim$10$^{-8}$ to $\sim$10$^{-11}$ for latest and earliest stellar types targeted by a space-based coronagraph. 
% where $a_\text{EEID} \propto R_\text{star} T_\text{star}^2$. 

Assuming that changes in the raw contrast in the images (i.e. speckle noise) is the limiting noise factor for planet detection, the wavefront error requirements will be proportional to $\sqrt{\epsilon}$. For the sake of simplicity, we assert that in order to image an exoplanet with $\epsilon = 10^{-10}$, the WFS needs to be sensitive to 1~pm in surface height. Thus, we can extrapolate the WFS surface height sensitivity requirement to the case of exoEarths around other stellar types by
\begin{equation}
    h_\text{req} \approx (1~\text{pm})\left(\frac{\text{1 au}}{a_\text{EEID}}\right).
\end{equation}

Focusing on the exoEarth science case, we can also set the spatial sampling requirements such that the WFS resolves the spatial frequencies that correspond to speckles within the field-of-view that encompasses the habitable zone (HZ)\cite{Kopparapu2013}. In other words, we can optionally select the number of pixels across the WFS image, $D_\text{pix}$, to optimize the contrast stability within an angular separation of 
\begin{equation}
    \alpha_\text{WFS} = \frac{D_\text{pix}}{2}\frac{\lambda}{D_\text{tel}}
\end{equation}
from the host star, where $D_\text{tel}$ is the effective telescope diameter and $\lambda/D_\text{tel}$ is angular resolution of the telescope. To ensure that the optimized field-of-view covers the full habitable zone and its immediate surroundings, we set $\alpha_\text{WFS}$ to twice the size of the outer edge HZ: $a_\text{HZ}/d_\text{star}$, where $a_\text{HZ}$ is the semi-major axis of the HZ outer boundary and $d_\text{star}$ is the distance to the star. This leads to the following sampling requirement: 
\begin{equation}
    D_\text{pix} = 4 \frac{a_\text{HZ}/d_\text{star}}{\lambda/D_\text{tel}}.
    \label{eqn:dpix}
\end{equation}
In the following, we consider WFS designs with either fixed sampling or variable sampling based on the stellar type.  

\begin{figure}[t!]
    \centering
    \includegraphics[width=0.5\linewidth]{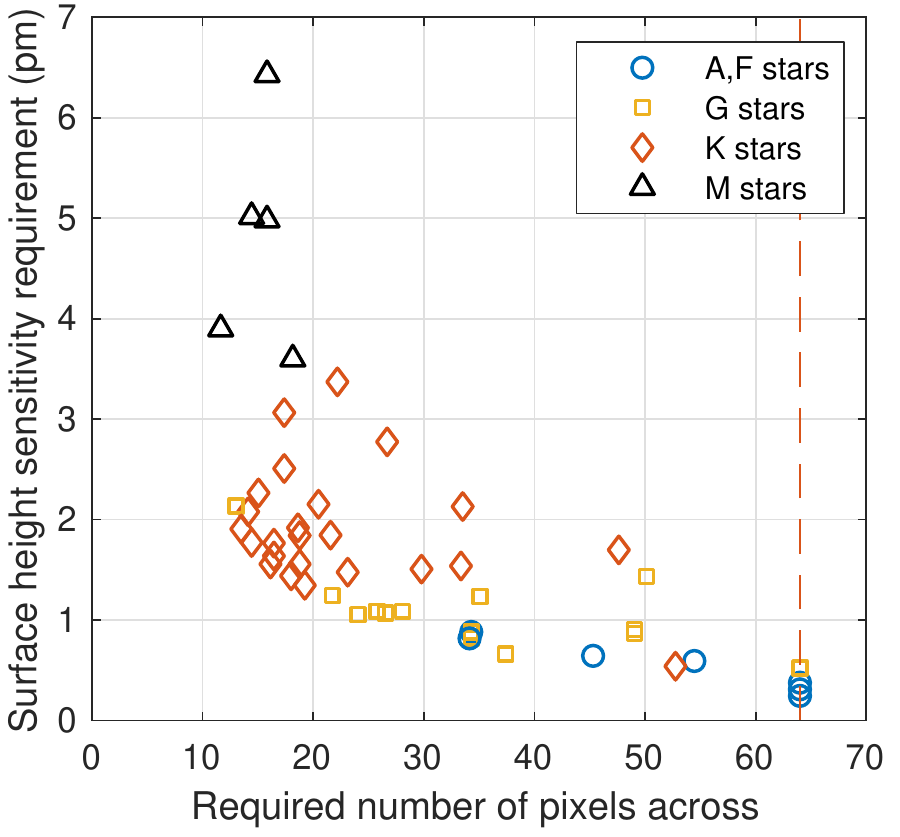}
    \caption{Requirements for imaging an Earth-sized exoplanet in the habitable zone of the 50 nearest HabEx targets. Late-type stars (K and M) have close-in habitable zones wherein an Earth-size planet has a favorable flux ratio and, therefore, the resolution and sensitivity requirements for the WFS are relaxed. Early type stars have a widely-separated habitable zone where a planet would have a more extreme flux ratio. A larger high-contrast field-of-view is achieved by using more pixels in the WFS image and the smaller planet-to-star flux ratios lead to tighter wavefront stability requirements. We limited the maximum number of pixels across the beam to 64 (indicated by the vertical dashed line).}
    \label{fig:habextargets_requirements}
\end{figure}

The HabEx report\cite{HabEx_finalReport} (Table D-1) provides a list of target stars used for the simulated exoplanet direct imaging survey. We use a representative sample for our analysis that includes the nearest 50 stars, which cover the full range of stellar types. Figure~\ref{fig:habextargets_requirements} shows the requirements ($h_\text{req}$ and $D_\text{pix}$) for each target based on the corresponding HZ separation. The surface height sensitivity requirement ranges from $h_\text{req}$~=~0.25~pm to $h_\text{req}$~=~6.5~pm for the most extreme cases, while the average requirement is $h_\text{req}$~=~1.8~pm. Since the HabEx DMs will have 64 actuators across, we limit the maximum number of samples to $D_\text{pix}$~=~64. The average $D_\text{pix}$ requirement is 30. For the case of larger telescopes such as LUVOIR, the $h_\text{req}$ requirements are the same, but $D_\text{pix} \propto D_\text{tel}$.

\subsection{Photon-noise-limited performance}

In this section, we present performance predictions assuming the ZWFS is photon-noise limited. Indeed, we demonstrated that the DST ZWFS noise is within a factor of 2 of the expected noise limit. Future missions will also likely have the advantage of photon-counting detectors (e.g. electron-multiplying CCDs), which makes the detector dark current and read noise negligible, especially after combining many WFS frames. Table~\ref{tab:HabExparams} shows the other relevant parameters for our analysis.  

\begin{table}[t]
    \centering
    \begin{tabular}{|c|c|c|c|}
        \hline
        Name & Symbol & Value \\\hline
        Telescope diameter & $D_\text{tel}$ & 4~m \\\hline
        Collecting Area & $A_\text{tel}^*$ & 12.6~m$^2$  \\\hline
        Transmission & $T$ & 0.3 \\\hline
        Quantum efficiency & $q$ & 0.9 \\\hline
        Dimple phase shift & $\theta$ & $\pi$/2 \\\hline
        Ref. wave intensity frac. & $f_\text{b}$ & 0.24 \\\hline
        WFS wave intensity frac. &$f_\text{Z}$ & 0.51 \\\hline
        Bandwidth & $\Delta\lambda/\lambda_0$ & 20\%\\\hline
    \end{tabular}
    \caption{Assumed parameters for HabEx error budgets. }
    \label{tab:HabExparams}
\end{table}

% \begin{table}[t]
%     \centering
%     \begin{tabular}{|c|c|c|c|}
%         \hline
%          & HabEx & LUVOIR B & LUVOIR A \\\hline
%         % Full Diameter & 4~m & 8.0~m & 15.0~m \\\hline
%         $D_\text{tel}^*$ & 4~m & 6.7~m & 13.5~m \\\hline
%         $A_\text{tel}^*$ & 12.6~m$^2$ & 35.3~m$^2$ & 142~m$^2$ \\\hline
%         $T$ & \multicolumn{3}{c|}{0.3}\\\hline
%         % Beam diameter & \multicolumn{3}{c|}{300~pixels}\\\hline
%         % Detector type & \multicolumn{3}{c|}{Zero noise} \\\hline
%         $q$ & \multicolumn{3}{c|}{0.9} \\\hline
%         % Read noise (rms) & \multicolumn{3}{c|}{0~e$^{-}$ (negligible)} \\\hline
%         % Dark current & \multicolumn{3}{c|}{3$\times$10$^{-5}$~e$^{-}$/pixel/sec } \\\hline
%         % Clock Induced Charge & \multicolumn{3}{c|}{2$\times$10$^{-3}$~e$^{-}$/pixel/frame}  \\\hline
%         % Pixel well depth & \multicolumn{3}{c|}{?~e$^{-}$} \\\hline
%         $\theta$ & \multicolumn{3}{c|}{$\pi$/2 }\\\hline
%         % $f_\text{A}$ & \multicolumn{3}{c|}{1.0}\\\hline
%         $f_\text{b}$ & \multicolumn{3}{c|}{0.24}\\\hline
%         $f_\text{Z}$ & \multicolumn{3}{c|}{0.51}\\\hline
%         $\Delta\lambda/\lambda_0$ & \multicolumn{3}{c|}{20\%}\\\hline
%     \end{tabular}
%     \caption{Assumed parameters for HabEx and LUVOIR error budgets. $^*$Using the inscribed diameter.  }
%     \label{tab:HabExLUVOIRparams}
% \end{table}

We characterize the WFS performance via the integration time needed to achieve a given sensitivity for each star in the target list via Eqns.~\ref{eqn:errortheory} and \ref{eqn:errortheory2}. Recalling that the error in the wavefront measurement is $\sigma_{\Delta\phi} = \beta_p / \sqrt{2 I_\text{e}}$, we assume $\chi^\prime(\phi_0)=1$ and $f_A = 1$, and solve for the WFS integration time: 
\begin{equation}
    \tau = \frac{f_\text{Z}}{2 f_b \dot{I}_\text{e} }\left(\frac{\lambda_0}{4\pi h_\text{req}}\right)^2,
    \label{eqn:tau}
\end{equation}
where $\dot{I}_\text{e} = I_\text{e}/\tau$ is the image count rate (electrons per second per pixel). The WFS loop update rate for the most critical range of spatial frequencies is approximately $\tau$. However, this requirement is based on to the worst-case spatial modes; since coronagraphs are designed to be robust to low-order aberrations\cite{Ruane2018_metrics}, they will have a relaxed $h_\text{req}$ (e.g. by a factor of $>$10$\times$ for vortex coronagraphs\cite{Ruane2018_JATIS}) and the wavefront correction may be updated at correspondingly faster rates. 

\subsubsection{NGS performance}

The performance in NGS mode depends on the brightness of the host star in the WFS passband. Table~\ref{tab:HabExLUVOIRbandselection} gives $\tau$ estimates for the HabEx telescope and coronagraph instrument. The 11~example targets were selected to span the full range of stellar types. We assume the WFS uses either a $B$, $V$, $R$, or $I$ filter and fixed pupil sampling (i.e. $D_\text{pix}$~=~64). The passbands have $\Delta\lambda/\lambda_0$~=~20\% and central wavelengths of 438~nm, 545~nm, 641~nm, and 798~nm, respectively. While the $B$ filter is optimal for an F star (e.g. Procyon), M stars require significantly longer integration times in $B$ compared to $I$. 

If only a single WFS passband is used for all 50 targets, the mean $\tau$ is 3.91~hr, 2.95~hr, 2.76~hr, and 3.54~hr in $B$, $V$, $R$, and $I$, respectively. Indeed, $R$ band is optimal for the majority (29 out of 50) of the targets considered. The capability to select between the WFS filters for each observation provides the best performance with a mean $\tau$ of 2.52~hr, and thus a potential 10\% improvement in mean WFS loop rate. The best combination of two filters is $B$ and $R$, which gives a mean $\tau$ of 2.62~hr. Using a subset of the desired filter options may have practical benefits, such as reducing the total number of optics and the wavelength range over which the optical coatings and detector quantum efficiency need to be optimized. In the following, we use the best option out of all four filters, but a subset of filters may provide a more practical solution without significantly impacting the performance in most cases. 

\begin{table}[t]
    \centering
    \begin{tabular}{|c|c|c|c|c|c|c|}
        \hline
        \textbf{Star Name} & \textbf{Type} & \textbf{Dist (pc)} & \multicolumn{4}{c|}{\textbf{Integration time (min)}} \\\hline
        & & & $B$ & $V$ & $R$ & $I$ \\\hline
        % Procyon & F5 & 3.5~pc & 0.8~min & 1.0~min & 1.1~min & 1.8~min \\\hline % index 6
        % eta Cas & G0 & 5.8~pc & 17~min & 17~min & 18~min & 25~min  \\\hline % index 16
        % tau Ceti$^*$ & G8 & 3.7~pc & 20~min & 18~min & 17~min & 21~min \\\hline% index 8
        % 82 Eri$^*$ & G8 & 6.0~pc & 40~min & 36~min & 34~min & 46~min \\\hline% index 19
        % sig Dra$^*$ & K0 & 5.8~pc & 62~min & 52~min & 49~min & 65~min \\\hline % index 15
        % 40 Eri$^*$ & K1 & 5.0~pc & 51~min & 42~min & 37~min & 47~min \\\hline% index 12
        % GJ 570 A$^*$ & K4 & 5.8~pc & 3.7~hr & 2.3~hr & 1.6~hr & 1.8~hr \\\hline % index 17 
        % eps Ind$^*$ & K5 & 3.6~pc & 81~min & 53~min & 51~min & 57~min \\\hline % index 9
        % 61 Cyg A$^*$ & K5 & 3.5~pc & 2.4~hr & 85~min & 56~min & 60~min \\\hline % index 5
        % 61 Cyg B$^*$ & K7 & 3.5~pc & 6.2~hr & 3.1~hr & 1.7~hr & 60~min \\\hline % index 4
        % AX~Mic & M1 & 4.0~pc & 12~hr & 5.6~hr & 5.0~hr & 3.5~hr  \\\hline % index 10
        Procyon & F5IV-V & 3.5 & 24.4 & 29.7 & 32.7 & 51.2  \\\hline
        eta Cas & G0V & 5.8 & 82.6 & 85.8 & 88.7 & 123  \\\hline
        tau Ceti$^*$ & G8V & 3.6 & 39.8 & 35.3 & 33.6 & 42.1  \\\hline
        82 Eri$^*$ & G8V & 6.0 & 108 & 96.6 & 91.9 & 123  \\\hline
        sig Dra$^*$ & K0V & 5.8 & 112 & 94.6 & 88.3 & 117  \\\hline
        40 Eri$^*$ & K1V & 5.0 & 89.0 & 72.8 & 64.3 & 82.1  \\\hline
        GJ 570 A$^*$ & K4V & 5.9 & 195 & 122 & 82.4 & 96.0  \\\hline
        eps Ind$^*$ & K5V & 3.6 & 73.6 & 48.2 & 45.9 & 52.0  \\\hline
        61 Cyg A$^*$ & K5V & 3.5 & 78.2 & 45.5 & 29.9 & 31.8  \\\hline
        61 Cyg B$^*$ & K7V & 3.5 & 134 & 67.4 & 37.6 & 21.7  \\\hline
        AX Mic & M1/M2V & 4.0 & 222 & 107 & 94.2 & 66.7  \\\hline
        %\multicolumn{3}{|l|}{Mean of 50 targets (up to 10~pc)}  & 5.7~hr & 2.9~hr & 2.3~hr & 2.2~hr & 1.9~hr \\\hline
        \multicolumn{3}{|l|}{Number of stars with lowest $\tau$}
         & 10/50 & 5/50 & 26/50 & 9/50 \\\hline

    \end{tabular}
    \caption{Estimated integration times using the primary starlight for wavefront sensing in $B$, $V$, $R$, and $I$ filters ($\lambda_0$ of 438~nm, 545~nm, 641~nm, and 798~nm, respectively.) with 64 pixels across the beam in the HabEx coronagraph instrument. The 11 stars listed are representative of the full range of spectral types. $^*$HabEx ``deep dive" targets. }
    \label{tab:HabExLUVOIRbandselection}
\end{table}

\begin{figure}[t!]
    \centering
    \includegraphics[width=\linewidth]{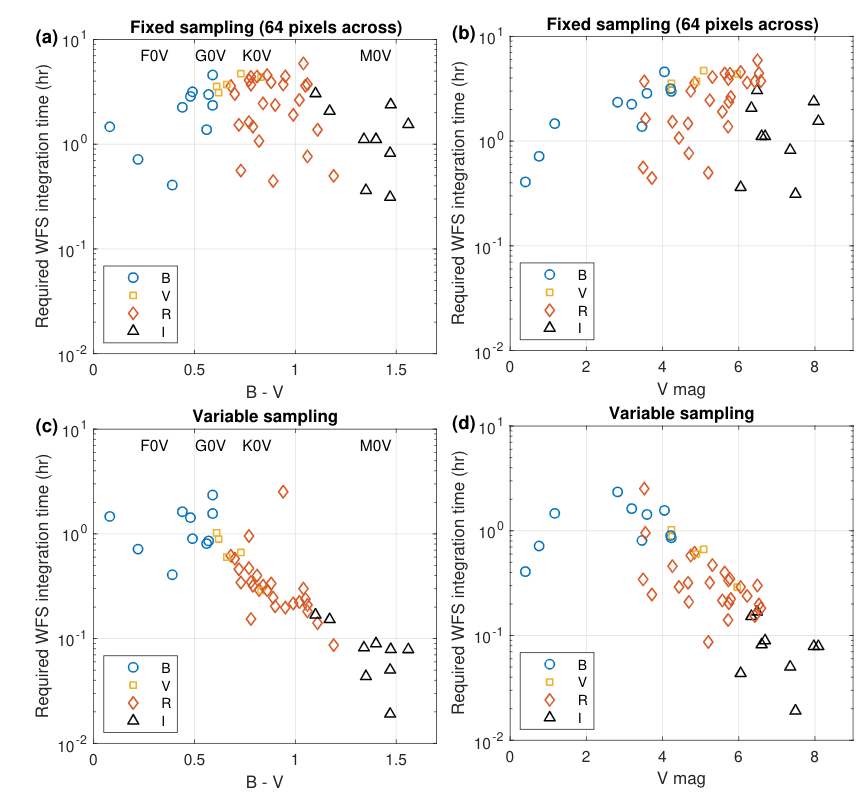}
    \caption{WFS integration time, $\tau$, needed to meet the surface height requirements, $h_\text{req}$, in Fig.~\ref{fig:habextargets_requirements} with the HabEx coronagraph instrument with (a)-(b)~$D_\text{pix}$~=~64 and (c)-(d) the $D_\text{pix}$ given in Fig.~\ref{fig:habextargets_requirements}. The columns shows the $B-V$ color and apparent $V$ magnitude of the host stars, respectively. The spectral types are noted along the top of the $B-V$ color plots. The marker shape indicates the optimal band for wavefront sensing for each target. These results can be readily scaled to the case of LUVOIR since $\tau \propto N_\text{pix}/A_\text{tel}$. In the case of variable sampling, $\tau$ is approximately constant for increasing $A_\text{tel}$.}
    \label{fig:habextargets_optimalband}
\end{figure}

Figure~\ref{fig:habextargets_optimalband} shows the $\tau$ estimates with both fixed and variable sampling using the optimal filter for each target. The fixed sampling case, as described above, has $D_\text{pix}$~=~64 for all observations. After optimizing the WFS passband for each star, the required $\tau$ is not correlated with the stellar type (Fig.~\ref{fig:habextargets_optimalband}a shows $B-V$ color) or apparent magnitude (Fig.~\ref{fig:habextargets_optimalband}b shows $V$ mag). However, there is a correlation with the stellar distance. The minimum $\tau$ is for the nearest stars; e.g. \textit{HD 95735} and \textit{61 Cyg B} have $\tau\approx$~20~min. The most distant stars included in our analysis are at $d_\text{star}$~=~10~pc and have $\tau\approx$~5~hr.

One way to reduce the $\tau$ requirement is to reduce the number of WFS pixels. Recall that Eqn.~\ref{eqn:dpix} gives the number of samples needed to stabilize the contrast in a field-of-view that extends to twice the HZ radius. Applying both the $h_\text{req}$ and $D_\text{pix}$ requirements shown in Fig.~\ref{fig:habextargets_requirements} leads to the $\tau$ in Fig.~\ref{fig:habextargets_optimalband}c,d. The required $\tau$ is drastically reduced compared to the fixed sampling case because coarser sampling is used in almost all cases. Late-type stars show the biggest improvement because the HZ is close-in and fewer WFS pixels are required to stabilize the wavefront at the most critical spatial frequencies. On the other hand, the brightest and earliest spectral types still have $D_\text{pix}$ of approximately 64~pixels, which leads to $\tau$ on the order of 1~hr.

The most challenging cases out of the nearest 50 HabEx targets are \textit{delta Eridani} and \textit{beta Hydri}, which are both subgiants that are significantly more luminous than a typical star in their spectral class (respectively, G and K). Despite the fact that they are among the brightest stars HabEx will observe, they have a widely separated HZ and, thus, an exoEarth orbiting them would have a relatively small $\epsilon$ and their observations would have the most extreme $h_\text{req}$ and $D_\text{pix}$ requirements. These two stars fall into a category of targets whose requirements are more difficult than a true Earth-Sun twin. Of the other 5 targets with $\tau>$~1~hr, the host stars are either A stars (i.e. \textit{Fomalhaut}), F stars (i.e. \textit{Pi$^3$ Orionis}, \textit{Gamma Leporis}), or early G type star with luminosities $>L_\odot$ (i.e. \textit{Beta Comae Berenices} and \textit{Iota Persei}, which have luminosities of 1.4~$L_\odot$ and 2.2~$L_\odot$, respectively). This luminous group of stars may ultimately drive the observatory stability requirements. However, it may be beneficial to treat stars with super-Solar luminosities as lower priority or only aim to detect planets significantly larger than Earth in their HZs. The result would make the observatory stability requirements safely $\sim$1~pm per hour using only the NGS mode. 

The LUVOIR coronagraph instrument benefits from a larger aperture (15~m for LUVOIR~A and 8~m for LUVOIR~B) because $\tau \propto N_\text{pix}/A_\text{tel}$ and the integration time needed for science observations is shorter in general. In the case where we allow variable sampling in the WFS image to match the same high-contrast field-of-view, $N_\text{pix} \propto D_\text{tel}^2$ and therefore the overall $\tau$ remains approximately constant with aperture diameter. Yet, the LUVOIR case is more challenging because the target list includes more distant stars and a larger fraction of the targets are early spectral types, especially F stars.

\subsubsection{Laser-based sensing}

Using a laser source instead of natural starlight is a way to drastically reduce the integration time needed to achieve a given $h_\text{req}$, but the ideal WFS configuration will depend on the telescope design. For instance, the segmented primary mirror in the case of LUVOIR will likely be prone to larger mid-spatial frequency aberrations than the monolith design of HabEx. Using the formation-flying LGS with LUVOIR has the potential to significantly relax the segment-to-segment phasing and stability requirements\cite{Douglas2019}. In the case of HabEx, designed to be stable to mid-spatial frequency aberrations, the capability to use NGS and an internal laser source may be a sufficient and cost-effective approach. Specifically, NGS light would be used for infrequent control of the wavefront along the full beam path, including the primary mirror aberrations, while the internal laser source would be used to compensate for errors that arise within the coronagraph instrument, especially due to unwanted changes in the DM surfaces. The internal source could potentially be installed near an image plane, upstream of the DMs, with an angular offset with respect to the optical axis allowing it to illuminate the coronagraph optics with out-of-band laser light. The laser light could then be reflected from an off-axis portion of the FPM with a ZWFS dimple, and allow for closed-loop control of the DM surface using a configuration similar to the DST ZWFS. 

With equivalent assumptions to the NGS case above, $D_\text{pix}$~=~64, and a 633~nm laser source, the integration time to achieve $h_\text{req}$~=~1~pm is $<$1~min if the equivalent source magnitude is $<$-1.7, which corresponds to $\dot{I}_\text{e}$~=~5$\times$10$^{7}$ counts per second in the WFS image. Other photon noise limited cases can be easily computed using Eqn.~\ref{eqn:tau}. 

\subsection{Error budgets including systematic errors}\label{sec:systematics}

The theoretical cases described above only considered the impact of photon noise on the differential wavefront measurements. In particular, we applied Eqn.~\ref{eqn:errortheory} to predict the impact of random errors. Eqn.~\ref{eqn:errortheory} also describes how any systematic uncertainty in the WFS image intensity leads to errors in the phase measurements. Regardless of the source of the uncertainty in the intensity, each image contributes an phase error of $\sigma_{I} (\Delta\phi  /  \Delta I)$, and the differential phase error is 
\begin{equation}
    \sigma_{\Delta\phi} = \sqrt{2}\frac{ \Delta\phi }{ \Delta I } \sigma_{I},
\end{equation}
where the factor of $\sqrt{2}$ accounts for the two images used in the differential measurement.

Here, we introduce four additional systematic error terms that may be significant in practice. The first is the pupil intensity calibration error, $\sigma_{I_\text{cal}}$, which causes a uncertainty in the phase of
\begin{equation}
    \sigma_{\Delta\phi} = \frac{\delta (\Delta\phi)}{\delta I_\text{cal}} \sigma_{I_\text{cal}} = \Delta\phi \frac{\sigma_{I_\text{cal}}}{2 I_\text{cal} },
\end{equation}
where $I_\text{cal}=A^2$. For example, if $\sigma_{I_\text{cal}}/I_\text{cal}$~=~10\%, the corresponding error in phase is 5\% of the actual phase. We refer to the factor of $1/2$ in this conversion as the error scale factor. 

Using a similar approach, an error in the knowledge of the reference wave amplitude, $\sigma_{b}$, corresponds to a phase uncertainty of 
\begin{equation}
    \sigma_{\Delta\phi} =  \Delta\phi \frac{\sigma_{b} }{b}.
\end{equation}
The error scale factor in this case is 1. However, an error in the model-based estimate of the reference wave could be related to the underlying assumption of the $\hat{d}$ parameter, which depends on wavelength ($\lambda_0$), focal ratio ($F^{\#}$), and dimple diameter ($d$). For reference, we compute an error of approximately 10\% in $b$ for an error of 6\% in $\hat{d}$ for the case where $\hat{d}=1$.

The next term we consider is the error in the dimple depth. For DST, we use the value provided by the manufacturer, which is $h_\text{Z}$~=~72$\pm$3~nm. For $\lambda_0$~=~575~nm, the phase shift is $\theta=4 \pi h_\text{Z}/\lambda_0=(0.50\pm0.02)\pi$. The resulting phase error is given by
\begin{equation}
    \sigma_{\Delta\phi} = \Delta\phi  \frac{\gamma}{\chi^\prime} \sigma_{\theta},
\end{equation}
where $\gamma = \cos(\phi_\text{0}) \cos\theta + \sin(\phi_\text{0})\sin\theta$ and $\chi^\prime = \cos(\phi_\text{0})\sin\theta + \sin(\phi_\text{0})(1-\cos\theta)$. The ratio $\gamma/\chi^\prime$ typically ranges from 0.3 to 0.7 for $\phi_0$ values of $\pm\pi/8$, but goes to infinity at the negative bound (1/4 wave). Conservatively assuming $\gamma/\chi^\prime \approx 1$, the dimple depth uncertainty on DST causes a 4\% error in the phase measurement. This uncertainty term should, in principle, also account for the error in the angle of incidence as well as the index of refraction for a transmissive ZWFS.

The last term captures the error in the knowledge of the nominal phase about which the WFS is making differential measurements, $\sigma_{\phi_0}$. The differential phase measurement has a resulting error of 
\begin{equation}
    \sigma_{\Delta\phi} = \Delta\phi \frac{\chi}{\chi^\prime}\sigma_{\phi_0},
\end{equation}
where $\chi = \sin(\phi_0) \sin\theta - \cos(\phi_0)(1-\cos\theta)$. The ratio $\chi/\chi^\prime$ is 1 for $\theta=\pi/2$ and $\phi_0=0$, but ranges from 0.4 to 2.4 for $\phi_0=\pm\pi/8$. For simplicity, we assume $\chi/\chi^\prime\approx1$ and set a requirement of $\sigma_{\phi_0}\le$~0.1~rad. This and the previous term will introduce errors that depend spatially on $\phi_0$. 
\begin{table}[t]
    \centering
    \begin{tabular}{|c|c|c|c|c|}
        \hline
        Error source & Error term & Scale factor & Requirement & Contribution \\\hline
        WFS images & $\sigma_{I}/\Delta I$ & $\sqrt{2}$ & 70\% & 0.99~pm \\\hline
        Pupil calibration & $\sigma_{I_\text{cal}}/I_\text{cal}$ & 1/2 & 10\% & 0.05~pm \\\hline
        Reference wave & $\sigma_{b}/b$ & 1 & 10\% & 0.1~pm \\\hline
        Dimple phase shift & $\sigma_{\theta}$ & $\gamma/\chi^\prime\approx 1$ & 0.1~rad & 0.1~pm \\\hline
        Initial phase & $\sigma_{\phi_0}$ & $\chi /\chi^\prime\approx 1$ & 0.1~rad & 0.1~pm \\\hline
        \multicolumn{4}{|r|}{\textbf{Total:}} & 1~pm\\\hline
    \end{tabular}
    \caption{An example error budget for picometer sensitivity including systematic effects. The total error budget adds the indiviudal contributions in quadrature following Eqn~\ref{eqn:errorbudget}.}
    \label{tab:systematics}
\end{table}

Combining all of the errors, the uncertainty in the differential phase measurement is given by
\begin{equation}
    \frac{ \sigma_{\Delta\phi} }{ \Delta\phi } = \sqrt{ 2 \left( \frac{\sigma_{I}}{\Delta I} \right)^2 + \left( \frac{\sigma_{I_\text{cal}}}{2 I_\text{cal}} \right)^2+ \left( \frac{\sigma_{b}}{b} \right)^2
    + \left( \frac{\gamma \sigma_{\theta}}{\chi^\prime} \right)^2 + \left( \frac{\chi \sigma_{\phi_0} }{\chi^\prime}\right)^2 }.
    \label{eqn:errorbudget}
\end{equation}
This expression is also the inverse of the $S/N$ ratio of the phase measurement. Under the assumptions outlined above, Table~\ref{tab:systematics} gives an example error budget for achieving 1~pm sensitivity including systematic errors. Our budget allocates 0.99~pm of error to noise in the WFS images to show that the laboratory demonstrations above should not have been significantly impacted by the systematic errors. The error budget allows for a 10\% error in the pupil calibration and reference wave model as well as a 0.1~rad error in the dimple phase shift and initial wavefront assumptions. It follows that the predictions for future space telescope based on the photon-noise-limited predictions are also robust to systematics. 

We did not include the dominant dynamic errors in this discussion, namely wavefront jitter and pupil shear with respect to the detector; future work will address the tolerances on these factors via end-to-end modelling with realistic wavefronts and other observatory dynamics. We also neglected systematic errors due to the wavelength dependence of the reconstructor, which are especially significant in the case of a broadband source where the source spectrum is unknown. In future work, we will develop a rigorous broadband reconstructor by generalizing the development in section \ref{sec:theory} to account for the wavelength dependence. However, for the sake of this demonstration, we emphasize that the error introduced by applying the simpler reconstructor derived for a single wavelength to the case of a broadband source is effectively similar to introducing uncertainties in $b$ and $\theta$, both of which have a relatively weak impact in the overall error budget (see Table~\ref{tab:systematics}). If necessary, these systematic errors can also be mitigated in practice through an empirical, source-specific calibration of the reconstructor.

\section{Conclusion} \label{sec:conc} 

We have presented a summary of the experimental demonstration of a ZWFS on the DST, validating the \textit{in situ} WFS concept baselined by both the HabEx and LUVOIR mission concepts in a similar optical configuration. Most significantly, we confirmed that the ZWFS is sensitive to the expected range of spatial frequencies, and the noise floor is within a factor of two of the theoretical prediction for picometer-level wavefront errors. In addition, we demonstrated closed-loop control resolving an individual DM actuator and, by extension, the ZWFS can be used to control spatial frequencies up to the limit set of the DM actuator spacing, which is critical for stabilizing high-contrast images. 

We also presented a detailed theoretical analysis of the predicted performance in the context of future space telescopes. Using natural guide stars, we show that the ZWFS will be able to sense picometer wavefront changes at $\sim$1~hr cadence for the worst-case stars. Laser illumination either with a external or internal source would reduce this time scale by a large margin; e.g. the equivalent of a magnitude -2 star theoretically gives picometer sensitivity in $<$1~min. Although our theoretical analysis focused on overcoming photon-noise, we also show that the systematic contributions to the error budget are expected to be relatively small. 

Future theoretical work will investigate the impact of dynamic errors, such as wavefront jitter and pupil shear, as well as systematic errors due to the wavelength dependence of the reconstructor in the case of broadband sources. Our coronagraph testbeds will continue to incorporate the development of wavefront sensing techniques and control algorithms. This, in combination with the in-flight demonstration planned by the Roman Space Telescope CGI, will quickly advance the WFS technology such that it readily meets the requirements of next generation of space telescopes with coronagraph instruments.

% \begin{itemize}
%     \item Validates the \textit{in situ} WFS concept used by both the HabEx and LUVOIR mission concepts in a similar optical configuration. 
%     \item demonstrated sensitivity within a factor of 2 of the theoretical prediction and closed loop control an individual actuators to demonstrate mid-spatial frequency sensitivity and operation.
%     \item We also presented a detailed theoretical analysis of the predicted performance in the context of future space telescopes. 
%     \item building upon WFIRST CGI 
% \end{itemize}

% \clearpage

\appendix

\section{Derivation}\label{sec:appendixDerivation}

\subsection{Zernike WFS image formation}

%We write the field in the pupil plane of interest as $E_\text{p}(x,y)=A_\text{p}(x,y)\exp(i\phi(x,y))$, where $A_\text{p}(x,y)$ is the pupil field amplitude and $\phi(x,y)$ is the phase error that we wish to estimate. 
The input pupil field, $E_\text{p}(x,y)$, is related to the field in the focal plane, $E_\text{f}(\xi,\eta)$, by a Fourier transform:
\begin{equation}
    E_\text{f}(\xi,\eta) = \text{FT}\left\{E_\text{p}(x,y)\right\},
\end{equation}
where 
\begin{equation}
    \text{FT}\{E(x,y)\} = \frac{1}{\lambda f}\iint E(x,y) e^{-ik(x\xi+y\eta)/f} dx dy,
\end{equation}
$k=2\pi/\lambda$, $\lambda$ is the wavelength, and $f$ is the focal length. The focal plane phase mask applies a phase shift, $\theta$, within a circular region with diameter, $d$, giving a complex transmission of
\begin{equation}
    t(\rho) = 1+\left[\left(e^{i\theta}-1\right)M\left(\frac{\rho}{d/2}\right)\right],
\end{equation}
where $\rho^2 = \xi^2 + \eta^2$ and 
\begin{equation}
    M(u) = 
    \begin{cases} 
      1 & u\leq 1 \\
      0 & u > 1 
   \end{cases}.
\end{equation}
After the mask, the focal plane field is $t(\rho)E_\text{f}(\xi,\eta)$  %$t(\rho)\;\text{FT}\left\{E_\text{p}(x,y)\right\}$ 
and the field in the subsequent pupil, where the WFS camera is located, is obtained via a second Fourier transform:
\begin{equation}
    E_\text{c}(x,y)% = \text{FT}\left\{E_\text{f}(\xi,\eta)\right\} 
    = \text{FT}\left\{t(\rho)\;\text{FT}\left\{E_\text{p}(x,y)\right\}\right\}.
\end{equation}
By the convolution theorem, the pupil field at the camera consists of an inverted image of the input pupil convolved with the Fourier transform of the mask function:
% \begin{equation}
%     E_\text{c}(x,y) = \text{FT}\left\{t(\rho)\right\}*\text{FT}\left\{\text{FT}\left\{E_\text{p}(x,y)\right\}\right\} = \text{FT}\left\{t(\rho)\right\}*E^\prime_\text{p}(x,y),
% \end{equation}
\begin{equation}
    E_\text{c}(x,y) = \text{FT}\left\{t(\rho)\right\}*E^\prime_\text{p}(x,y),
\end{equation}
where $E^\prime_\text{p}(x,y)=E_\text{p}(-x,-y)$. The convolution kernel may be written 
\begin{equation}
    \text{FT}\left\{t(\rho)\right\} = \delta(x,y) + \left(e^{i\theta}-1\right)\text{FT}\left\{M\left(\frac{\rho}{d/2}\right)\right\},
\end{equation}
where $\delta(x,y)$ is the Dirac delta function and
\begin{equation}
\text{FT}\left\{M\left(\frac{\rho}{d/2}\right)\right\} = \frac{k d^2}{4f}\frac{J_1\left(k_r\right)}{k_r}%= \frac{\pi d^2}{2\lambda f}\frac{J_1\left( k d \rho/2f\right)}{k d \rho/2f},
\end{equation}
is an Airy diffraction pattern with $k_r = k d r/2f$ and $r^2 = x^2 + y^2$. Thus, the pupil field at the camera is
% \begin{equation}
%     E_\text{c}(x,y) = E_\text{p}(-x,-y) + \left(e^{i\theta}-1\right)\left[\frac{k d^2}{4f}\frac{J_1\left(k_r\right)}{k_r}*E_\text{p}(-x,-y)\right].
% \end{equation}
% \begin{equation}
%     E_\text{c}(x,y) = A^\prime_\text{p}(x,y)e^{i\phi^\prime(x,y)} + \left(e^{i\theta}-1\right)b(x,y), 
% \end{equation}
\begin{equation}
    E_\text{c}(x,y) = \left[E^\prime_\text{p}(x,y) - b(x,y)\right] + b(x,y)e^{i\theta}, 
    \label{eqn:interference}
\end{equation}
where 
\begin{equation}
    b(x,y) = \frac{k d^2}{4f}\frac{J_1\left(k_r\right)}{k_r}*E^\prime_\text{p}(x,y).
\end{equation}
Equation~\ref{eqn:interference} illustrates that the phase mask takes the light in the core of the focal spot and shifts the phase by $\theta$ such that it interferes with the rest of the beam at the camera. For this reason $b(x,y)$ is also known as the reference wave. 

For simplicity, we denote the pupil field as $E^\prime_\text{p}(x,y)=A(x,y)\exp(i\phi(x,y))$, where $A(x,y)$ is the field amplitude and $\phi(x,y)$ is the phase we wish to estimate. The image on the camera is given by $I_\text{Z}(x,y) = |E_\text{c}(x,y)|^2$, which we expand as\cite{NDiaye2013_ZELDA}
\begin{equation}
    I_\text{Z} = A^2 + 2 b^2(1-\cos\theta) + 2 A b \chi,
    \label{eqn:IZapp}
\end{equation}
where $\chi = \sin\phi \sin\theta - \cos\phi (1-\cos\theta)$. Herein, and in the main text, it is implicit that the intensity images, $A$, $b$, and $\phi$ are two-dimensional functions and we omit the $(x,y)$ coordinates. 

\subsection{Phase reconstruction}

Solving Eqn.~\ref{eqn:IZapp} for $\phi$ is a means to estimate the input pupil phase from two images: $I_\text{Z}$ and $I_0$, which are respectively the images with and without the focal plane mask. In practice, $I_0$ is usually obtained by offsetting the the focal plane mask by several times $d$ such that the phase dimple has a negligible impact on the image. The pupil field amplitude is determined by $A=\sqrt{I_0}$. 

\subsubsection{The reference wave}

The field for the reference wave, $b$, is estimated via a model of the optical system, which is most often calculated as 
\begin{equation}
    b = \text{FT}\left\{M\left(\frac{\rho}{d/2}\right)\text{FT}\left\{E_\text{p}(x,y)\right\}\right\}
\end{equation}
using an approximate version of $E_\text{p}(x,y)$. 
For a circular pupil with amplitude $A$ and diameter $D$, the field in the focal plane is an Airy diffraction pattern:
\begin{equation}
    \text{FT}\left\{E_\text{p}\right\} = A \frac{k D^2}{4f}\frac{J_1\left(k_\rho\right)}{k_\rho},
\end{equation}
where $k_\rho = k D \rho/2f$ and due to the rotational symmetry: 
\begin{equation}
    b = \frac{2\pi}{\lambda f}\int_0^{d/2} \text{FT}\left\{E_\text{p}\right\} J_0(k \rho r /f ) \rho d\rho.
\end{equation}
Recalling that the second order expansion of the $J_0$ Bessel function about $r=0$ is $J_0(x)\approx1-x^2/4$, integrating each term in the series individually yields a simple analytical approximation to estimate the reference wave profile:
\begin{equation}
    b = A\left[ 1- J_0\left(\pi \hat{d}/2\right)- \left(\pi \hat{d}/2\right)^2 J_2\left(\pi \hat{d}/2\right) \left(\frac{r}{D}\right)^2 \right],
    \label{eqn:bmodelapp}
\end{equation}
where $r<D/2$, $\hat{d}=d/(\lambda F^\#)$, and $F^\#=f/D$. The approximation is valid if $r\ll\lambda f/d$; i.e., where the Airy pattern in the pupil plane is significantly larger than the pupil itself. $\hat{d}$ is effectively the ratio between the dimple diameter and the PSF size. For example, the values of $b$ at the center ($r=0$) and edge ($r=D/2$) of the pupil are approximately 0.53 and 0.37, respectively, for $\hat{d}$~=~1. Based on Eqn.~\ref{eqn:bmodelapp}, the mean value of $b$ over the pupil is 
\begin{equation}
    \bar{b} = A\left[ 1- J_0\left(\pi \hat{d}/2\right)- \frac{1}{8}\left(\pi \hat{d}/2\right)^2 J_2\left(\pi \hat{d}/2\right) \right],
    % \label{eqn:bmodel}
\end{equation}
which is typically accurate to within a few percent for $\hat{d} \approx 1$. %We find that $\bar{b}$ is particularly useful for use in a ZWFS error budget. 

\subsubsection{Taylor expansions for phase approximations}

The estimation of $\phi$ can be simplified significantly in the regime where $\phi\ll$~1~rad. Assuming $\sin\phi\approx\phi$ and $\cos\phi\approx1$,  
\begin{equation}
    \phi = \frac{1}{\sin\theta}\left( \frac{I_\text{Z}}{2Ab} - \frac{A}{2b} + \left( 1-\frac{b}{A} \right) \left( 1-\cos\theta \right) \right).
\end{equation}
Using instead the second order Taylor expansion about $\phi=0$, $\sin\phi\approx\phi$ and $\cos\phi\approx1-\phi^2/2$,  
\begin{equation}
    \phi = \frac{\sqrt{c} - b A \sin\theta}{Ab(1-\cos\theta)},
\end{equation}
where
\begin{equation}
    c = 2Ab\sin(\theta/2)^2\left[I_\text{Z}-A^2-2b^2+3Ab+b(2b-A)\cos\theta\right].
\end{equation}

% a1 = 2*b.*P.*(-2*b.^2+IZ+3*b.*P-P.^2+b.*(2*b-P)*cos(theta))*sin(theta/2)^2;
% a2 = b.*P.*sin(theta);
% numer = sqrt(a1) + a2;
% denom = b.*P*(cos(theta)-1);
% phz = -1*real(numer./(denom+1e-30)); 

\subsubsection{Full analytical solution}

For the application of space-based high-contrast imaging, the phase is not necessarily small, but the required precision is extremely small. Therefore, in order to minimize systematic errors in our phase measurements, we use an exact analytical solution obtained using the Mathematica\cite{mathematica} Solve function:
\begin{equation}
    \phi = \pm \arccos\left\{\frac{c_1\pm\sqrt{c_2}}{4 b^2 A^2(\cos\theta-1)}\right\},
\end{equation}
where
\begin{equation}
    c_1 = A b (I_\text{Z}-A^2)(1-\cos\theta)-2 A b^3(1-\cos\theta)^2,
\end{equation}
\begin{equation}
    c_2 = b^2 A^2 \sin^2\theta (c_3 - c_4 - c_5 - c_6),
\end{equation}
\begin{equation}
    c_3 = 4b^2(I_\text{Z}+A^2),
\end{equation}
\begin{equation}
    c_4 = (I_\text{Z}-A^2)^2,
\end{equation}
\begin{equation}
    c_5 = 4 b^2(I_\text{Z}+A^2-2 b^2)\cos\theta,
\end{equation}
\begin{equation}
    c_6 = 2b^4(3+\cos(2\theta)).
\end{equation}

\acknowledgments     
We thank Stuart Shaklan and Stefan Martin for useful discussions. Keith Patterson led the design and commissioning of the DST. This work was carried out at the Jet Propulsion Laboratory, California Institute of Technology, under contract with the National Aeronautics and Space Administration.

%%%%%%%%%%%%%%%%%%%%%%%%%%%%%%%%%%%%%%%%%%%%%%%%%%%%%%%%%%%%%
%%%%% References %%%%%

\bibliography{refLibrary}   %>>>> bibliography data in report.bib
\bibliographystyle{spiebib}   %>>>> makes bibtex use spiebib.bst

\end{document}